\documentclass[fleqn,12pt]{wlscirep}
\usepackage[utf8]{inputenc}
\usepackage[T1]{fontenc}
\usepackage{bm}
\usepackage{setspace}
\usepackage{soul}
\sethlcolor{yellow}
\usepackage{lineno}
\usepackage{chemformula}
\usepackage{amsmath}
\usepackage{amssymb}
\usepackage{siunitx}
\usepackage{multirow}
\usepackage[normalem]{ulem}
\usepackage{amsmath,amssymb}
\usepackage{cases}

\title{Inertial Dynamics of a Skymeron}

\author[1,15,\textbf{*}]{Mona Bhukta}
\author[1]{Duc Minh Tran}
\author[1]{Kilian Leutner}
\author[1,2]{Takaaki Dohi}
\author[3,\textbf{*}]{Nikolai S. Kiselev}
\author[4]{Filipp N. Rybakov}
\author[4,5]{Olle Eriksson}
\author[1]{Fabian Kammerbauer}
\author[6]{Sebastian Wintz}
\author[6]{Markus Weigand}
\author[7,8,9]{Hendrik Ohldag}
\author[1,10]{Maria-Andromachi Syskaki}
\author[1]{Robin Tietgen}
\author[1]{Edoardo Mangini}
\author[1]{Sabrina Kerber}
\author[11]{Thibaud Denneulin}
\author[11]{Joseph Vimal Vas}
\author[11]{Rafal E. Dunin-Borkowski}
\author[12]{Bastian Pfau}
\author[13]{Oleg A. Tretiakov}
\author[1,\textbf{*}]{Robert Frömter}
\author[1, 14,\textbf{*}]{Mathias Kläui}

\affil[1]{Institute of Physics, Johannes Gutenberg University Mainz, 55099 Mainz, Germany}
\affil[2]{Laboratory for Nanoelectronics and Spintronics, Research Institute of Electrical Communication, Tohoku University, 2-1-1 Katahira, Aoba, Sendai 980-8577, Japan}
\affil[3]{Peter Grünberg Institute, Forschungszentrum Jülich and JARA, 52425 Jülich, Germany}
\affil[4]{Department of Physics and Astronomy, Uppsala University, Box-516, Uppsala SE-751 20, Sweden}

\affil[5]{Wallenberg Initiative Materials Science, WISE, Uppsala University, Box-516, Uppsala SE-751 20, Sweden}
\affil[6]{Helmholtz-Zentrum Berlin für Materialien und Energie GmbH, Hahn-Meitner-Platz 1, 14109 Berlin,
Germany}
\affil[7]{Advanced Light Source, Lawrence Berkeley National Laboratory, 6 Cyclotron Rd, Berkeley, CA, 94720 USA}
\affil[8]{Department of Physics, University of California Santa Cruz, Santa Cruz USA}
\affil[9]{Department of Materials Science and Engineering, Stanford University, Stanford USA}
\affil[10]{Singulus Technologies AG, Hanauer Landstrasse 107, 63796 Kahl am Main, Germany}
\affil[11]{Ernst Ruska-Centre for Microscopy and Spectroscopy with Electrons, Forschungszentrum Jülich, 52425 Jülich, Germany}
\affil[12]{Max Born Institute for Nonlinear Optics and Short Pulse Spectroscopy, 12489 Berlin, Germany}
\affil[13]{School of Physics, The University of New South Wales, Sydney, New South Wales 2052, Australia}

\affil[14]{Center for Quantum Spintronics, Department of Physics, Norwegian University of Science and Technology, 7491 Trondheim, Norway}
\affil[15]{Université Grenoble Alpes, CNRS, CEA, SPINTEC, 38054 Grenoble, France}

\affil[*]{mbhukta@uni-mainz.de, n.kiselev@fz-juelich.de, froemter@uni-mainz.de, klaeui@uni-mainz.de}

\begin{abstract}

Topological spin textures in antiferromagnets inherit the compensated magnetic order of the host material, resulting for skyrmions in the suppression of the skyrmion Hall effect and enabling ultrafast dynamics that make them attractive for low-power spintronic devices. In intrinsic antiferromagnets, the two sublattices are identical, but synthetic antiferromagnets offer additional control mechanisms by enabling independent tuning of the properties of two ferromagnetic layers. Analogous to a ferrimagnet, the total magnetic moment can be adjusted by compensating the two-layer moments. As a unique, additional tuning possibility, one can adjust all other magnetic parameters independently for each layer. Here, we demonstrate that tuning the effective magnetic anisotropy difference between the two layers drives a spin-flop transition where one layer reorients and thus stabilizes an orthogonal configuration with one layer oriented along the out-of-plane direction and the other in-plane. In this transition, an antiferromagnetic skyrmion undergoes a homotopic reconfiguration into a complex spin texture comprising a skyrmion coupled to an in-plane bimeron. Element-specific X-ray microscopy resolves this texture layer by layer: one hosts an out-of-plane skyrmion coupled to an in-plane bimeron in the other. We refer to this previously unexplored three-dimensional spin texture as a skymeron. By using time-resolved pump–probe X-ray microscopy, we discover unique polarity-dependent dynamics during current pulses: a short post-pulse inertia-like continuation of the motion, and a slower return towards the pinned initial state. Micromagnetic simulations reveal that the inertia-like propagation originates from the finite-time relaxation of reorientation of the skymeron. Our results establish layer-selective anisotropy engineering as a route to uncharted composite spin textures with internal dynamical degrees of freedom, not possible in conventional antiferromagnets with identical sublattices.

\end{abstract}

\begin{document}
\doublespacing
%\flushbottom
%\setstretch{2.2}
\maketitle

\thispagestyle{empty}
\section*{Introduction}

Nanoscale chiral magnetic spin textures, including magnetic skyrmions\cite{
fert2013skyrmions, jiang2015blowing}, bimerons\cite{jani2021antiferromagnetic,bhukta2024homochiral}, have emerged as potential candidates for mobile bits in next-generation non-volatile memory \cite{tomasello2014strategy, zhang2015magnetic} and as synaptic and neuronal elements in neuromorphic \cite{song2020skyrmion,da2025neuromorphic} and unconventional computing \cite{raab2022brownian, beneke2024gesture} architectures. Their size, down to the nanometre regime, and electrical controllability by spin-orbit torques (SOTs) \cite{woo2016observation,jiang2017direct,litzius2017skyrmion} make them promising for fast, low-power operation. The current-driven dynamics of such spin textures are theoretically described  by the Thiele equation \cite{thiele1973}, obtained by applying a rigid-body approximation to the first-order Landau--Lifshitz--Gilbert (LLG) equation and reducing the continuous magnetization field to the translational coordinate of a quasiparticle. In this description, topology is encoded by a single integer-valued invariant, the winding number \cite{BP1975,nagaosa2013topological}, while the absence of an effective mass term precludes inertial dynamics. Nevertheless, an inertia-like motion emerges as a signature of non-rigidity, attributed either to topology coupled to the breathing mode in individual skyrmions\cite{buttner2015dynamics} or to pinning-mediated collective deformation of a skyrmion lattice in the creep regime\cite{birch2024dynamic}. However, owing to their large stray fields, limited thermal stability, and intrinsic skyrmion Hall effect (SkHE) \cite{jiang2017direct,litzius2017skyrmion}, ferromagnetic (FM) materials remain fundamentally constrained, motivating a shift towards magnetic spin textures in antiferromagnets (AFMs). There, cancellation of the net gyrovector at angular-momentum compensation suppresses the SkHE \cite{barker2016static,zhang2016magnetic,zhang2016antiferromagnetic,dohi2019formation}, while negligible stray fields enable dense skyrmion packing and SOT-driven velocities approaching 1 km/s \cite{pham2024fast}. Although potentially ultrafast transient inertia is predicted for skyrmions in AFM\cite{panigrahy2022skyrmion}, the timescales are too fast so that no measurable inertial response has been observed even on sub-nanosecond timescales \cite{bhukta2025antiferromagnetic}, and the dynamics remain captured by the single-particle description, with the winding number defined on the Néel vector rather than the magnetization \cite{barker2016static}.

Extending magnetic order into the third dimension (3D) \cite{fernandez2017three} or across multiple sublattices \cite{kent2021creation,grelier2022three} could increase information-storage density and enable biologically inspired functionality through magnetic networks that emulate synaptic connectivity \cite{zhang2023magnon}. Although considerable efforts have focused on the realization of topological spin textures beyond two-dimensional cases, including hopfions \cite{zheng2023hopfion,chen2026_hopfion_rings}, cocoons \cite{grelier2022three} and chiral bobbers \cite{zheng2018experimental}, their response to electrical currents, which is essential for their controlled electrical manipulation, remains largely unexplored. A key challenge is to understand whether these dynamics involve internal degrees of freedom, such as relative rotations and non-rigid spin structure deformations, associated with their more complex topology~\cite{Rybakov2025} and underlying magnetic configuration~\cite{Bimerons2025}, and therefore not captured by the conventional rigid-texture Thiele description~\cite{thiele1973}. Such modes can enlarge the dynamical state space of a texture, while their finite-time relaxation can provide fading memory, both potentially useful for temporal information processing in reservoir ~\cite{raab2022brownian} and other unconventional computing approaches~\cite{everschor2024topological}. Discovering such dynamics requires, firstly, complex non-collinear alignment of sub-lattices that enable complex spin textures. And secondly, one needs sub-nanosecond time-resolved microscopy, as quasi-static imaging cannot access transient non-rigid deformation~\cite{rhensius2010imaging}, inertia-like behaviour~\cite{Doering1948}, or other dynamics beyond the rigid-particle picture in magnetic spin textures extending beyond two dimensions and in multilayers with individually tuned sublayer properties.

%Extending magnetic order into the third dimension (3D)~\cite{fernandez2017three} or across multiple sublattices~\cite{kent2021creation,grelier2022three} could increase information-storage density and enable biologically inspired functionality through magnetic networks~\cite{zhang2023magnon}. However, electrical control of such complex spin textures remains largely unexplored, despite the realization of hopfions~\cite{zheng2023hopfion,chen2026_hopfion_rings}, cocoons~\cite{grelier2022three} and chiral bobbers~\cite{zheng2018experimental}. A key challenge is to determine whether their current-driven dynamics involve internal degrees of freedom, such as relative rotations and non-rigid deformations, that arise from their composite magnetic structure and richer topology~\cite{Rybakov2025,Bimerons2025} and are not captured by the conventional rigid-texture Thiele description~\cite{thiele1973}. Resolving such dynamics requires time-resolved microscopy, as quasi-static imaging cannot access transient non-rigid deformation~\cite{rhensius2010imaging}, inertia-like behaviour~\cite{Doering1948}, or other dynamics beyond the rigid-particle picture in magnetic excitations extending beyond two dimensions and in multilayers with individually tuned sublayer properties.

In this work, we experimentally realize the skymeron, a three-dimensional composite non-collinear spin texture, formed by coupled skyrmion and bimeron spin textures whose description extends beyond the single winding number of a conventional skyrmion, and resolve its current-induced dynamics on nanosecond timescales. To achieve this, we exploit synthetic antiferromagnets (SyAFMs), consisting of ferromagnetic layers separated by non-magnetic spacers and coupled through interlayer exchange \cite{parkin1990oscillations}, which provide an experimentally accessible platform for stabilizing homochiral antiferromagnetic spin textures such as skyrmions \cite{dohi2019formation,dohi2022enhanced,pham2024fast} and bimerons \cite{bhukta2024homochiral}. Here, we independently tune the effective anisotropy $K_{\mathrm{eff}}$ of the two ferromagnetic layers, a degree of control not available in intrinsic AFMs with symmetry-equivalent sublattices. The sign of $K_{\mathrm{eff}}$ selects the preferred polar or planar character. Engineering opposite signs of $K_{\mathrm{eff}}$ in the two layers enables an out-of-plane (OOP) field-driven asymmetric spin-flop transition, where one layer remains predominantly OOP magnetized while the other becomes in-plane (IP) magnetized. The system can thereby host unconventional spin textures comprising an OOP skyrmion in one layer and an IP bimeron in the other. During this transition, an AFM skyrmion continuously transforms into a coupled OOP skyrmion--IP bimeron spin texture, which we term a \textit{skymeron}, reflecting its dual character and intermediate position in the homotopy classification between skyrmions and merons. Using layer-resolved scanning transmission X-ray microscopy (STXM), we directly image the skymeron in real space. Time-resolved pump--probe X-ray microscopy with sub-nanosecond temporal resolution reveals that the distinct skyrmion and bimeron constituents remain coupled under SOT excitation. Due to the complex topology, they exhibit polarity-dependent non-rigid dynamics, including transient elongation and an inertia-like post-pulse continuation followed by slower relaxation. Micromagnetic simulations show that this response originates from the finite-time relaxation of skymerons, enabling complex dynamics not found in conventional spin textures.

\section*{Topological Spin Textures in Synthetic Antiferromagnets}

To describe spin textures, particularly useful are topological invariants, such as winding numbers, which distinguish different magnetic spin textures and enter their effective equations of motion, thereby shaping their responses to external driving \cite{buttner2015dynamics}. Therefore, we start by analysing the topology of possible composite spin textures in our tunable synthetic AFMs.
The topological classification depends on the order-parameter space and on the boundary conditions imposed by the magnetic background. 
In a two-dimensional ferromagnet~\cite{BP1975}, the order parameter takes values on $\mathbb{S}^2$, and for a collinear OOP background, the relevant homotopy group is $\pi_2(\mathbb{S}^2)=\mathbb{Z}$, leading to the integer skyrmion number, $Q\in \mathbb{Z}$.
When the effective anisotropy favors an easy-plane configuration, the boundary magnetization is no longer fixed to a single point but belongs to a different subspace of~$\mathbb{S}^2$. 
In this case, the classification changes qualitatively~\cite{Rybakov2025}, and the topological state is characterized by two integers from the free abelian group~$\mathbb{Z}\times\mathbb{Z}\equiv\mathbb{Z}^2$, referred to as meron indices~\cite{Rybakov2025, Bimerons2025, chen2026_hopfion_rings}.

In our SyAFM, the situation becomes richer because the order parameter is defined on the product space $\mathbb{S}^2 \times \mathbb{S}^2$, corresponding to the two coupled layers. 
Depending on the anisotropy and magnetic field, different boundary conditions yield distinct topological classes of localized spin textures (see Methods for details).
For strong perpendicular anisotropy and a low applied magnetic field, both layers are (near-) collinear, forming a compensated antiferromagnetic background. 
In this regime, which we call the \textit{polar-polar regime}, localized textures are characterized by a pair of integers associated with the two layers and correspond to antiferromagnetic skyrmions, schematically shown in Fig.~\ref{fig:1}\textbf{a}.
The topological charge of such a spin texture is defined by an ordered pair of integers \cite{barker2016static,bessarab2019stability}, which defines the skyrmion charge in layers A and B, respectively, $(Q^{\mathrm{(A)}}, Q^{\mathrm{(B)}})=(-1,-1)$.
The background magnetization surrounding such skyrmions in this case is nearly orthogonal to the plane of the multilayer, as depicted in Fig.~\ref{fig:1}\textbf{d}.
If the effective anisotropy favors the easy-plane regime in both layers, antiferromagnetic bimerons \cite{shen2020current,li2020bimeron} can be stabilized (Fig.~\ref{fig:1}\textbf{b}).
In this case, the background magnetization is nearly in-plane as schematically shown in  Fig.~\ref{fig:1}\textbf{e}.
The most intriguing situation arises at intermediate anisotropies, where the two layers experience different effective boundary conditions. 
As the magnetic field increases, the equilibrium magnetization of one layer is driven toward the film plane, while the other remains OOP magnetized. 
Here, we refer to this configuration as \textit{polar-planar} or nearly orthogonal regime, meaning that the net magnetizations in the two layers tend to be perpendicular to each other.
Note that a similar polar-planar regime may emerge in the case of ferromagnetic interlayer exchange coupling or van der Waals interactions~\cite{goerzen2026emergence}. In the following sections, we show that, in SyAFMs, this regime can be achieved through a novel mechanism: by combining anisotropy engineering with an external field.    

% In the following sections, we show that by tuning the anisotropy, a nearly orthogonal state can be achieved even in the absence of an applied field.

%
Changes in boundary conditions in a nearly orthogonal regime alter the underlying topological classification, enabling composite spin textures that combine the properties of skyrmions and bimerons.
The corresponding homotopy group, $\mathbb{Z}^3$,  consists of triples of integers. 
In this case, the composite spin texture can be thought of as one layer hosting a skyrmion, while the other layer hosts a bimeron \cite{gobel2019magnetic, zarzuela2020stability}. 
A schematic representation of such a composite spin texture is shown in Fig.~\ref{fig:1}\textbf{c}, and its background magnetization is schematically depicted in Fig.~\ref{fig:1}\textbf{f}.
Because such a spin texture simultaneously exhibits characteristics of both skyrmion and bimeron, we refer to it as a \textit{skymeron}.
Contrary to antiferromagnetic skyrmions, the topological index of this configuration is defined by a triplet of integers, $(q^\text{(A)}_\text{b}, q^\text{(B)}_\text{t}, q^\text{(B)}_\text{b}) = (-1, -1, 1)$ (see Methods for details). By judiciously applying a magnetic field, we can induce the transition between the polar-planar and polar-polar boundary conditions, which then leads to a natural mapping between the corresponding topological classes and transforms an AFM skyrmion into a skymeron (see Methods). In the following, we explore the experimental realization of the skymeron.
%
%This reflects the continuous transformation of a skymeron into an antiferromagnetic skyrmion (see Methods).
%
%In the following, we explore the experimental realization of such composite textures in SyAFM multilayers, where a skyrmion-like structure in one sublattice is coupled to a bimeron in the other, forming the composite skymeron quasi-particle.

%

%Unlike intrinsic antiferromagnets, SyAFMs allow independent engineering of the magnetic energy landscape in each ferromagnetic sublattice. Previous work has mainly utilized this flexibility to tune magnetic compensation \cite{hirata2019vanishing,caretta2018fast,woo2018current,dohi2019formation,dohi2022enhanced,dohi2024observation}, producing ferrimagnet-like behaviour and enabling efficient current-driven skyrmion dynamics. Here we exploit an additional degree of freedom: sublattice-specific control of the effective magnetic anisotropy. The resulting anisotropy imbalance breaks the symmetry between the coupled layers and enables the stabilization of new classes of topological spin textures.

\section*{Experimental Realization of Skymerons}

To experimentally check if skyrmerons can be realized and if their complex topology really leads to the unusual dynamics expected, we need to realize skymerons by deliberately breaking the symmetry between the two antiferromagnetically aligned layers of a SyAFM. By engineering effective anisotropies of opposite sign in the two FM layers, an applied out-of-plane field drives the system into the mixed polar–planar regime that hosts the skymeron.

\begin{figure}[h!]
    \centering
    \includegraphics[width = 18cm]{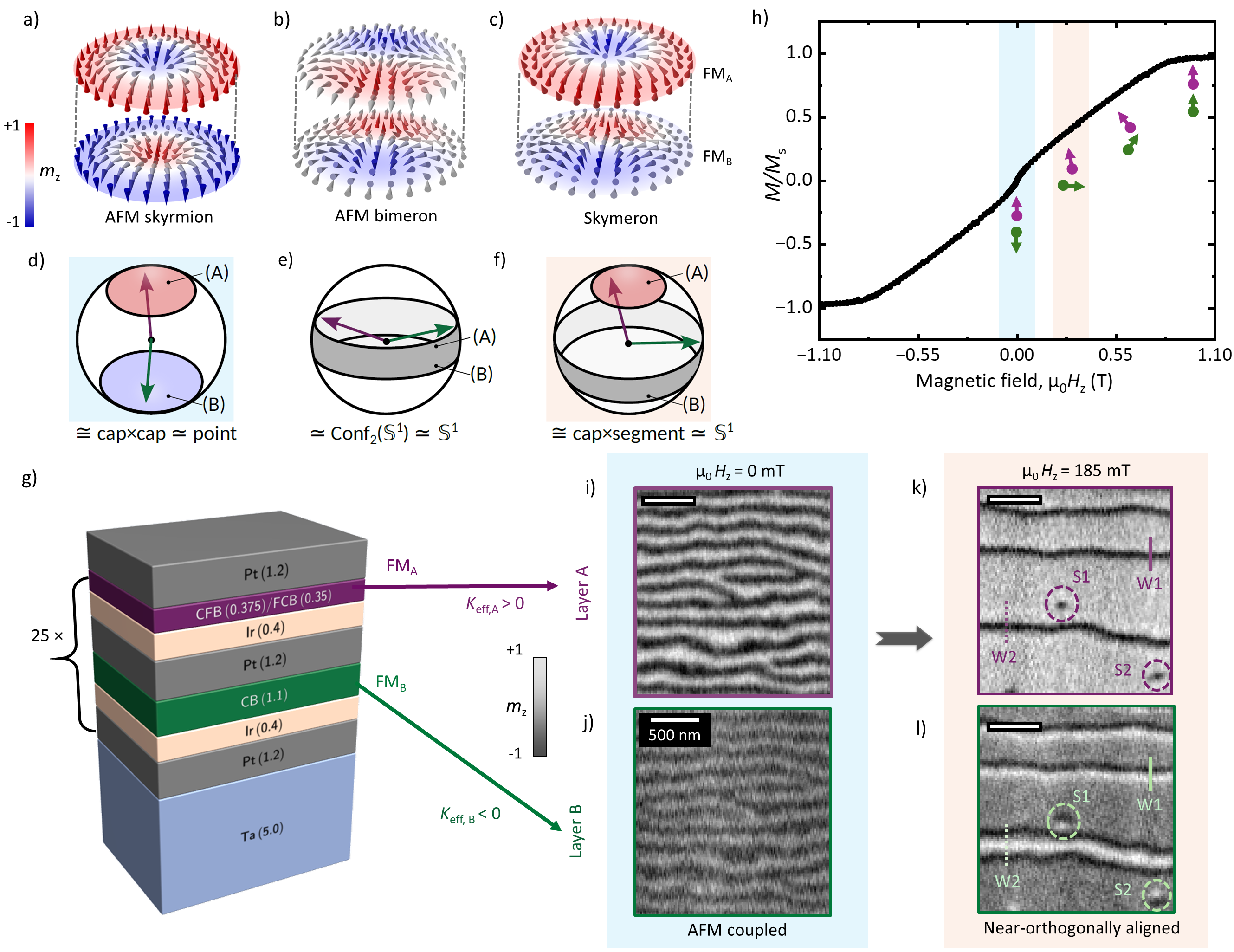}
      \captionsetup{width=0.99\textwidth}
        \caption{\textbf{Experimental realization of skymerons in a synthetic antiferromagnet.}
        \textbf{(a-c)} Schematics of an AFM skyrmion, an AFM bimeron, and a skymeron, respectively, in SyAFM. The skymeron consists of a skyrmion-like spin texture in layer A coupled to a bimeron in layer B. \textbf{(d)}, \textbf{(e)}, and \textbf{(f)} Schematics of the subspaces corresponding to the background configurations of the composite spin textures shown in (a), (b), and (c), respectively (see Table~\ref{table_glossary} for a description of terms). 
        These sketches illustrate the subsets of the order-parameter space that minimize the energy density of the homogeneous state.
        %
        %These spaces also include moderate angular deviations to account for imperfections.
        %
       \textbf{(g)} Schematic of the SyAFM stack with thicknesses in nm. \textbf{(h)} Normalized OOP magnetic hysteresis loop, measured by SQUID. \textbf{(i–l),} XMCD–STXM imaging of the magnetic configuration at the Fe $L_3$ and Co $L_3$ absorption edges. Opposite contrast at zero field confirms antiparallel OOP magnetization of the two FM layers (\textbf{i, j}). In the near-orthogonal regime, element-selective imaging reveals a skyrmion in layer A (\textbf{k}) and a bimeron in layer B (\textbf{l}), demonstrating the formation of the skymeron.
        }
        \label{fig:1}
\end{figure}

To achieve this, we design and optimize a low-pinning SyAFM stack (see Methods) consisting of 25 repetitions of two alternating ferromagnetic layers A and B, illustrated in Fig.~\ref{fig:1}(g).
As in our previous study\cite{bhukta2025antiferromagnetic}, layer A is Fe-rich and layer B is Co-rich, so element-resolved X-ray magnetic circular dichroism (XMCD)–STXM at the Fe and Co $L_3$ edges provides element-selective, layer-specific contrast. Layer~A is engineered with positive effective anisotropy, $K_{\mathrm{eff,A}}>0$, favouring out-of-plane alignment and skyrmion formation \cite{dohi2024observation, bhukta2025antiferromagnetic}, while layer~B is tuned into the easy-plane regime, $K_{\mathrm{eff,B}}<0$, favouring bimerons \cite{bhukta2024homochiral}. At zero field, however, the antiferromagnetic interlayer exchange dominates over this anisotropy imbalance and enforces an antiparallel out-of-plane alignment of the two FM layers. As a result, the multilayer forms a multidomain state, reflected by the vanishing remanence in the magnetic hysteresis loop (Fig.~\ref{fig:1}(h)) and by the anti-aligned domains observed at the Fe and Co $L_3$ absorption edges (Fig.~\ref{fig:1}(i,j)).

%Sublattice A is engineered to exhibit perpendicular magnetic anisotropy ($K_{\mathrm{eff,A}} > 0$), favoring skyrmions, whereas sublattice B is tuned toward easy-plane anisotropy ($K_{\mathrm{eff,B}} < 0$), favoring bimerons. Under ambient conditions, the antiferromagnetic interlayer exchange exceeds the effective anisotropy energy ($E_{\mathrm{interlayer}}>E_{\mathrm{anisotropy}}$), enforcing collinear alignment of the two sublattices. Consequently, the multilayer exhibits an effective OOP multidomain configuration governed by the positive net anisotropy. This collective magnetic state is reflected by the vanishing remanence in the SQUID hysteresis loop (Fig.~\ref{fig:1}(h)) and by the multidomain pattern observed in XMCD–STXM imaging. Measurements at the Fe $L_3$ and Co $L_3$ absorption edges (Fig.~\ref{fig:1}(i,j)) reveal opposite magnetic contrast in the two sublattices at $\mu_0H_\mathrm{z}=0$ mT, directly confirming antiparallel OOP magnetization and antiferromagnetic interlayer coupling.

Upon increasing the out-of-plane magnetic field, the Zeeman energy competes with the anisotropy and interlayer exchange energies, driving the multilayer towards a spin-flop reorientation~\cite{bloemen1994oscillatory}. Because the two FM layers are energetically inequivalent, the easy-plane layer cants much more strongly than the layer retaining out-of-plane magnetization, converting the spin-flop, which is symmetric in conventional AFMs, into an asymmetric canting process (see Supplementary Note 4). At $\mu_0 H_\mathrm{z}=185~\mathrm{mT}$, the system enters a near-orthogonal regime in which the magnetization of one layer remains predominantly OOP, whereas that of the other becomes IP. Layer-resolved XMCD--STXM then reveals spatially co-localized but distinct textures in the two layers (Fig.~\ref{fig:1}(k,l)). At the Fe $L_3$ edge (layer A), the localized reversal of $m_\mathrm{z}$ identifies a skyrmion core, whereas the paired black--white lobes at the Co $L_3$ edge (layer B) are the characteristic XMCD signature of a bimeron in an in-plane background, arising from meron and antimeron cores of opposite polarity~\cite{bhukta2024homochiral}. The one-to-one spatial correlation between the skyrmion-like core in layer A and the bimeron in layer B identifies the bound composite spin texture as a skymeron, as sketched in Fig.~\ref{fig:1}(c). Although the normal-incidence XMCD contrast identifies the expected constituent textures, assignment of the complete \(\mathbb Z^{3}\) topological indices requires additional analysis.

\begin{figure}[h!]
    \centering
    \includegraphics[width = 17.5cm]{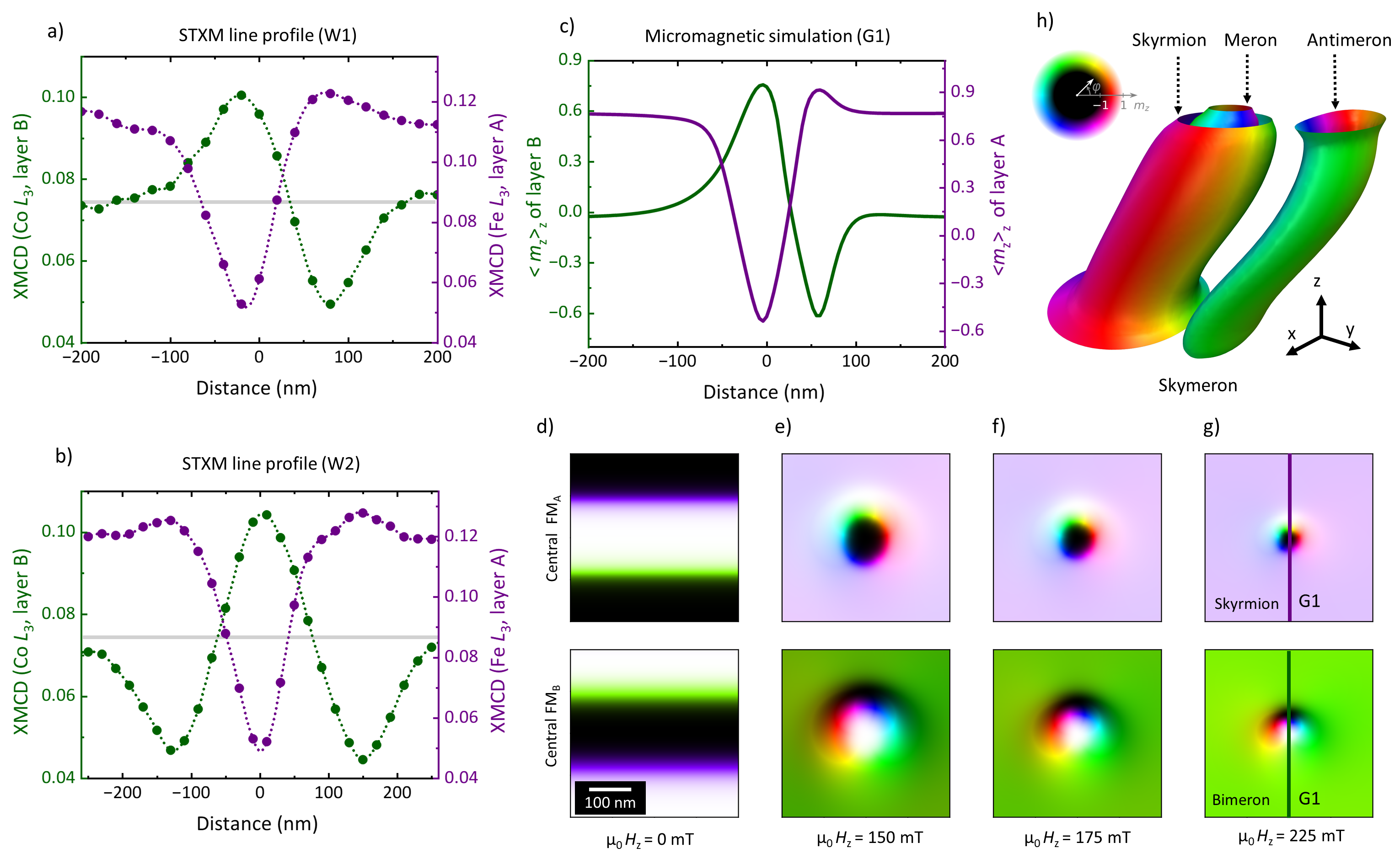}
      \captionsetup{width=0.99\textwidth}
\caption{\textbf{ Field-driven evolution and three-dimensional spin texture of the skymeron.} \textbf{(a,b)}, XMCD--STXM line profiles measured along W1 and W2 across the 360° and 540° domain walls, respectively (see Fig.~\ref{fig:1}(k,l)). The magenta and green curves correspond to the signals from layers A (Fe edge) and B (Co edge), respectively. \textbf{(c)}, Thickness-averaged micromagnetic line profiles of layers A and B of the skymeron shown in g). \textbf{(d--g)}, Magnetization maps extracted from the two central ferromagnetic layers A and B, showing the simulated field evolution of the skymeron. At zero field (\textbf{d}), the system stabilizes antiferromagnetically aligned stripe domains. At $\mu_0 H_z = 150\,\mathrm{mT}$ (\textbf{e}), isolated skymerons emerge. With a further increase of the field (\textbf{f,g}), the relative orientation of the two layer backgrounds progressively approaches a near-orthogonal configuration while the topological cores remain anti-aligned. \textbf{(h)} Three-dimensional contour rendering of the simulated skymeron at $\mu_0 H_z = 225\,\mathrm{mT}$. The isosurfaces correspond to constant out-of-plane magnetizations $m_\mathrm{z}=\cos\theta$. For the skyrmion layer, the isoline is shown at $\theta=90^\circ$, while for the bimeron layer the isolines correspond to $\theta=40^\circ$ and $\theta=140^\circ$.}
    \label{fig:2}
\end{figure}

To quantify the internal magnetization spin texture of the skymerons, we analyze XMCD line profiles along the directions $W_1$ and $W_2$ indicated in Fig.~\ref{fig:1}(k,l). The profiles along $W_1$ are shown in Fig.~\ref{fig:2}(a), where the magenta and green curves correspond to layers~A and B, respectively. In layer~A, the profile deviates from the symmetric shape expected for an ideal N\'eel skyrmion~\cite{boulle2016room,moreau2016additive}, revealing a pronounced anisotropic distortion that breaks mirror symmetry. In layer~B, by contrast, the profile exhibits the characteristic two-lobed line shape of a meron--antimeron pair, as observed in the SyAFM platform in prior work\cite{bhukta2024homochiral}. The profile along \(W_2\) (shown in Fig.~\ref{fig:2}(b)) further confirms the reversal of the surrounding in-plane background magnetization across the stripe (see supplementary Note 3.1).

Micromagnetic simulations reproduce the asymmetric skyrmion profile in layer A and the two-lobed bimeron signature in layer B (Fig.~\ref{fig:2}(c)), with thickness-averaged profiles in agreement with the STXM data. With increasing field, the system evolves from AFM-coupled stripe domains at zero field to isolated skymerons at $\mu_0H_\mathrm{z}=150~\mathrm{mT}$ and subsequently towards a near-orthogonal alignment of the two FM layers (Fig.~\ref{fig:2}(d-g); Supplementary Notes 4 and 5). Unlike in ferromagnetic multilayers, where an out-of-plane field primarily rescales the skyrmion radius and ultimately drives its collapse~\cite{buttner2018theory}, the field along the easy axis (OOP) in the present SyAFM tunes the relative layer orientation without destroying the composite topological cores. The evolution between the snapshots in Fig.~\ref{fig:2}(e-g) is continuous in our simulations and involves no creation or annihilation of topological defects, indicating the real-space manifestation of the homotopy path from the polar–planar \(\mathbb Z^{3}\) sector to the polar–polar \(\mathbb Z^{2}\) sector. Simulations of the full 25-bilayer stack further reveal that the skymeron is intrinsically three-dimensional, with a depth-dependent deformation rather than a straight texture through the film thickness (Fig.~\ref{fig:2}(h)). Although this spin texture is not directly resolved by the thickness-integrated STXM signal, it can account for the apparent elongation in Fig.~\ref{fig:1}(l). Taken together, these results establish the skymeron experimentally as a composite topological spin texture directly resolved in real space. The broken symmetry arising from the inequivalent FM layers introduces additional internal degrees of freedom in line with the complex topology and going beyond the rigid-body description of compensated AFM skyrmions, leading to the unconventional dynamics discussed below.

\section*{Current-Driven Skymeron Dynamics}

Conventional antiferromagnetic skyrmions translate as rigid quasiparticles under damping-like SOT (DL-SOT), with transient inertial dynamics confined to picosecond timescales by the strong interlayer exchange \cite{panigrahy2022skyrmion, bhukta2025antiferromagnetic}. The skymeron, by contrast, carries an additional internal degree of freedom: the meron–antimeron axis of the bimeron component breaks the rotational symmetry~\cite{shen2020current, vorobyev2024asymmetric}, enabling more complex current-induced dynamics. This allows SOTs to drive rotational and deformation modes alongside translation, producing a more complex and potentially non-rigid dynamical response. To resolve this response on its intrinsic nanosecond timescale, we employ time-resolved pump–probe STXM (see Methods), which provides the sub-nanosecond temporal resolution required to access transient processes such as non-rigid motion and possible inertial effects that are inaccessible to static post-excitation measurements. 

%Because STXM measurements are performed at a single photon energy at a time, we record the dynamics of the two magnetic layers sequentially at the Fe and Co $L_3$ absorption edges.

We probe skymeron dynamics using alternating current pulses with duration $10\, \mathrm{ns}$ and current density $J_{\mathrm{x}}=\pm1.3\times10^{12}\,\mathrm{A\,m^{-2}}$ (see the pulse waveform in Supplementary Fig.~S13). The alternating pulse polarity is employed with the aim of driving the skymerons back and forth along the track \cite{litzius2017skyrmion, bhukta2025antiferromagnetic} by means of the SOT. Owing to its asymmetric internal magnetization and composite topology, however, the skymeron responds differently to the two current polarities, which is qualitatively different from skyrmions \cite{litzius2017skyrmion, bhukta2025antiferromagnetic}. Consequently, alternating pulses alone do not restore its initial configuration, as shown by micromagnetic simulations in the next section. Resolving this unconventional response by pump--probe imaging therefore requires billions of excitation cycles to begin from the same well-defined magnetic configuration and follow a reproducible trajectory; stochastic or non-recurrent events would otherwise be averaged out, resulting in a loss of time-resolved magnetic contrast. In our experiment, we realize this reproducibility by making use of the naturally occurring, non-flat energy landscape, whose local energy minima define stable reference positions and restoring pathways, combined with an appropriately chosen current density. By tuning the current density, we can generate sufficient torques to induce dynamics, while the torques are low enough that the system returns to its original state. 

\begin{figure}[h!]
    \centering
    \includegraphics[width = 17cm]{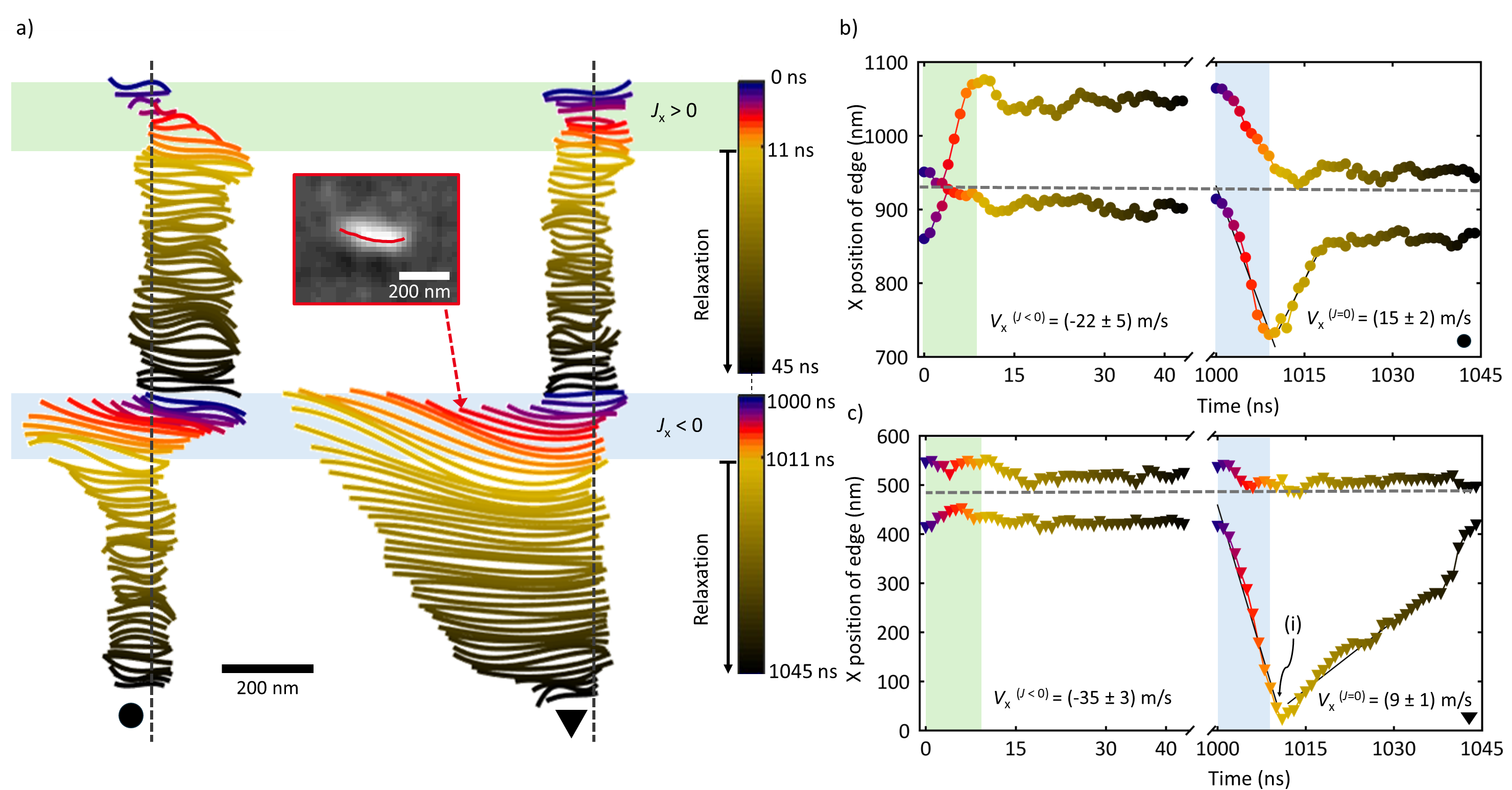}
      \captionsetup{width=0.99\textwidth}
\caption{\textbf{Spatiotemporal evolution of skymeron dynamics during a cycle of bipolar current pulses.} \textbf{(a)} Time-resolved contour evolution of two representative skymerons under bipolar current excitation. The contours are extracted from XMCD–STXM images at the Fe $L_3$ edge and vertically offset to visualize the temporal evolution during the excitation cycle. The color scale encodes time. Green (0–10 ns) and blue (1000–1010 ns) shaded regions denote intervals of positive and negative current polarity ($J_x>0$ and $J_x<0$), respectively. Black vertical dashed lines mark the pinning positions that anchor the skymerons during the motion. 
\textbf{(b,c)} Temporal evolution of the left and right $x$-edge positions for the skymerons marked by a circle (b) and a downward triangle (c) in (a). When the $J_x<0$ pulse is switched off, two successive processes occur: an inertia-like continuation of the motion (shown in (i)), followed by a slower relaxation toward the equilibrium position. Linear fits yield a current-driven velocity $v_x^{(J<0)} = (-35 \pm 3)\,\mathrm{m\,s^{-1}}$ and an average relaxation velocity $v_x^{J=0} = (9 \pm 1)\,\mathrm{m\,s^{-1}}$.}
    \label{fig:3}
\end{figure}

Fig.~\ref{fig:3}(a) shows the experimental time-resolved evolution of two representative skymerons imaged at the Fe $L_3$ edge, which provides XMCD contrast from their out-of-plane skyrmion cores. From each image, the skyrmion-core boundary is extracted and reduced to a centreline that tracks the temporal evolution of the texture shape (see Methods). Successive centrelines are vertically offset for visualization, while the green ($0$--$10$ ns) and blue ($1000$--$1010$ ns) regions denote the positive- and negative-current pulses, respectively. The two representative skymerons, marked by a circle and a downward-pointing triangle, are tracked throughout the pump--probe cycle.

%Fig.~\ref{fig:3}(a) shows the experimental temporal evolution of two representative skymerons imaged at the Fe $L_3$ absorption edge, which provides XMCD contrast from their out-of-plane skyrmion cores. At each time point, the skyrmion core was segmented from the corresponding XMCD image and its closed boundary was extracted. At each longitudinal position, the transverse coordinates of the upper and lower boundary branches were averaged to obtain the one-dimensional core centreline shown in Figure ~\ref{fig:3}(a). The resulting centreline was spatially smoothed to suppress pixel-scale variations. Successive centrelines were vertically offset and stacked to visualize their evolution throughout the excitation sequence, with colour encoding time. The green ($0$--$10$ ns) and blue ($1000$--$1010$ ns) shaded regions mark the positive and negative current pulses, respectively, whereas the subsequent traces show the corresponding post-pulse evolution. The two skymerons, denoted by a circle and a downward-pointing triangle, are tracked throughout the pump-probe cycle.

The time-resolved dynamics reveal behaviours that are drastically different from those observed for ferromagnetic \cite{litzius2017skyrmion} and antiferromagnetic skyrmions \cite{bhukta2025antiferromagnetic}. We can categorize the response into two regimes: dynamics during the current pulse and inertia-like propagation/relaxation after the pulse. First, during the current pulse, the skymerons depart from rigid-body motion. Their leading and trailing edges advance sequentially rather than simultaneously, a motion we describe as \textit{caterpillar-like} to distinguish it from the rigid-body translation of conventional antiferromagnetic skyrmions, in which both edges advance together. For instance, for the skymeron marked by the circle in Fig.~\ref{fig:3}(a), the left edge initially retracts while the right edge subsequently advances after 4 ns during the $J_\mathrm{x}>0$ pulse. However, the topology does not prescribe this caterpillar-like pathway for free motion; the observed dynamics can originate from the different energy landscapes experienced by the left and right edges of each skymeron.

%However, this sequential edge motion is not a consequence of the topology of the skymeron, as can be concluded from the different motion of the observed skymerons. Instead, it most probably originates from the different pinning energy landscapes experienced by the left and right edges of each skymeron.
%However, the topology does not prescribe this caterpillar-like pathway; it most probably originates from the different pinning energy landscapes experienced by the left and right edges of each skyrmion. Instead, topology only constrains the skyrmion and meron-type constituents to persist while their relative positions and shape evolve within the same topological sector.

A signature of the skymeron dynamics is its strong asymmetry under current reversal. For $J_\mathrm{x}>0$, the skymeron exhibits only a small net displacement, whereas $J_\mathrm{x}<0$ produces pronounced elongation, as quantified by the $x$-edge trajectories in Figs.~\ref{fig:3}(b,c). During the $J_\mathrm{x}<0$ pulse, the texture additionally develops a transient local rotation and curvature along $y$ at the left edge (Fig.~\ref{fig:3}(a)). The corresponding $y$-edge trajectories in Extended Data Fig.~2(a,b) quantify this transverse response and reveal a transient reorientation of the meron--antimeron axis relative to the current-flow direction ($x$). The driven response therefore involves internal deformation and rotation of the spin structure rather than rigid translation alone, while remaining within the same topological sector. The polarity dependence of this non-rigid response, leading to asymmetric dynamics for the two polarities, originates from the broken rotational symmetry of the underlying spin texture. As shown by the micromagnetic simulations in the next section, this symmetry breaking renders the SOT-driven dynamics intrinsically asymmetric under current reversal.

%A signature of the skymeron dynamics is the fact that the dynamics is strongly asymmetric under current reversal. For $J_\mathrm{x}>0$, the skymeron exhibits only a comparatively small net displacement, whereas for $J_\mathrm{x}<0$ the texture undergoes pronounced elongation. This behavior is quantified in Figs.~\ref{fig:3}(b,c), which show the temporal evolution of the left and right $x$-edge positions for the two representative skymerons. In each panel, the upper (lower) trace corresponds to the left (right) edge of the texture. Importantly, during the applied pulse for $J_\mathrm{x}<0$, the skymeron develops a transient local rotation and curvature along $y$ at the left edge, as visible in Fig.~\ref{fig:3}(a)

Second, after the current pulse is switched off, the skymerons exhibit a unique two-stage post-pulse response comprising an inertia-like forward continuation followed by a slower return towards their initial configurations. Upon removal of the drive, the skymeron continues to move for approximately 2 ns, producing an additional forward displacement of $\sim 20$--$60$ nm, as indicated by region (i) in Fig.~\ref{fig:3}(c). This forward continuation is followed by a slower back-relaxation, during which the right edge retracts by approximately $270~\mathrm{nm}$ towards the equilibrium configuration over a timescale of $\sim 30 $ ns. The two stages occur in opposite directions: the initial post-pulse motion follows the direction of the preceding current-driven displacement, whereas the later relaxation brings the texture back towards its equilibrium position. This separation of timescales and directions shows that the initial inertia-like motion cannot be explained solely by the restoring force of the local energy landscape. Instead, it indicates a relaxation of the excess energy stored in the current-induced deformation, resulting, for instance, in an increase in the micromagnetic anisotropy energy. This deformation remains coupled to the translational coordinate after removal of the drive and gives rise to the inertia-like response. Importantly, the pronounced transient elongation developed by the skymeron during current-driven motion indicates a non-rigid deformation in which the key micromagnetic energy contributions, including the anisotropy, magnetostatic, Zeeman, and interlayer-exchange energies, are redistributed, storing excess energy that is released upon pulse termination and acts as an energy reservoir for the subsequent inertia-like motion.

%Following the short inertia-like forward motion, the subsequent recovery of the deformed topological excitation provides a measure of the slower pinning-mediated relaxation. As shown in Fig.~\ref{fig:3}(c), the right edge retracts by approximately $270$ $\mathrm{nm}$ over a timescale of about $30$ $\mathrm{ns}$. This retraction after the current pulse is switched off is attributed to the restoring force of the local effective pinning potential together with relaxation of the internal magnetic energy stored during skymeron elongation, dominated by the excess anisotropy energy and accompanied by redistribution of the magnetostatic, Zeeman, and interlayer-exchange contributions. 

Following the short inertia-like forward motion, the subsequent recovery of the deformed skymeron reflects a slower relaxation driven by the restoring force of the local energy landscape, together with relaxation of the internal magnetic energy stored due to the elongation of the skymeron. A linear fit yields a relaxation velocity of $v_\mathrm{x}^{J=0} = (9 \pm 1)\,\mathrm{m\,s^{-1}}$, nearly four times smaller than the current-driven velocity $v_\mathrm{x}^{(J_\mathrm{x}<0)} = (-35 \pm 3)\,\mathrm{m\,s^{-1}}$. This reduction indicates that the local restoring force drives the motion at an effective scale of approximately one-quarter of the current-driven one. In the transverse direction, the current-driven and relaxation velocities are comparable, $v_\mathrm{y}^{(J_\mathrm{x}<0)}=(-11\pm1)\,\mathrm{m\,s^{-1}}$ and $v_\mathrm{y}^{J=0}=(10\pm1)\,\mathrm{m\,s^{-1}}$, respectively. The similar magnitudes of $v_\mathrm{x}^{J=0}$ and $v_\mathrm{y}^{J=0}$ indicate that the relaxation is not strongly anisotropic within the film plane. While the current-driven velocity  $v_\mathrm{x}^{(J<0)}$ is largely universal (see Supplementary Note 6 for other examples), the back-relaxation is governed by the local non-flat energy landscape and therefore yields skymeron-dependent relaxation velocities $v_\mathrm{x}^{J=0}$ and $v_\mathrm{y}^{J=0}$. The two post-pulse stages therefore have distinct physical origins. The initial \(1\)–\(2\) ns continuation can be attributed to the release of internal magnetic energy stored in the current-induced deformation, whereas the restoring force of the non-flat energy landscape dominates the later reverse motion.

A key signature of the skymeron response is its asymmetry under current reversal, such that alternating pulses alone do not restore an identical configuration after each cycle. Reproducible pump--probe imaging is nevertheless enabled by local energy minima, indicated by the black dashed lines in Fig.~\ref{fig:3}(a), which confine each skymeron and provide a well-defined reference configuration over billions of excitation cycles. The detailed deformation amplitudes and relaxation pathways vary with the local non-flat energy landscape, as observed for the circle- and triangle-marked textures. Nonetheless, this non-rigid, inertia-like, and asymmetric response is observed universally across skymerons and tracks, independent of current polarity, whose microscopic origin is elucidated in the next section using micromagnetic simulations.

\section*{Micromagnetic Simulations of Skymeron Dynamics}

\begin{figure}[h!]
    \centering
    \includegraphics[width = 17.5cm]{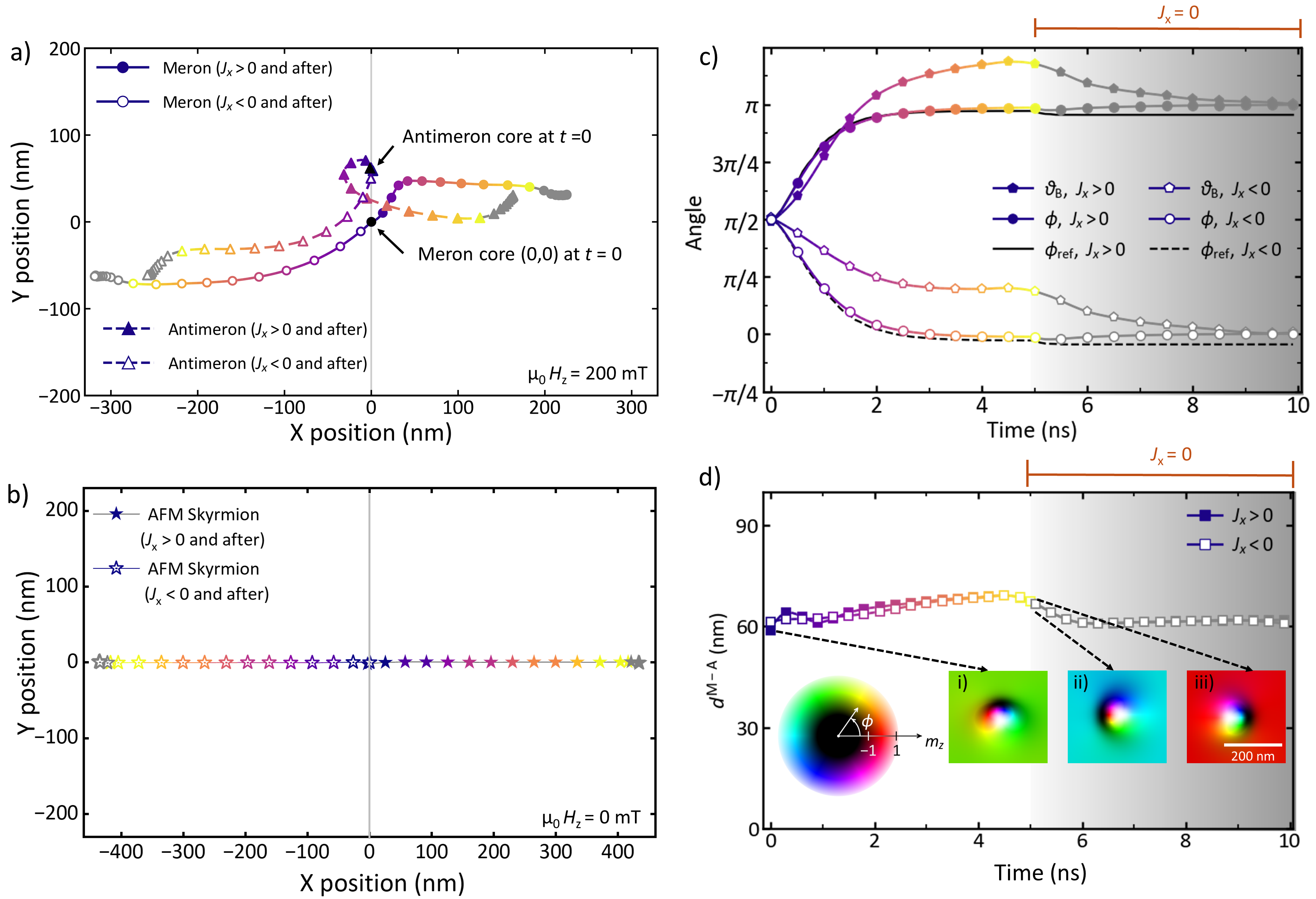}
      \captionsetup{width=1\textwidth}
\caption{\textbf{Micromagnetic simulations of non-rigid and inertia-like skymeron dynamics.} \textbf{(a)} Simulated current-driven trajectories of a skymeron stabilized at $\mu_0H_z=200$ $\mathrm{mT}$ for a 5~ns current pulse. Circles and triangles track the meron and antimeron cores, respectively. Filled symbols correspond to $J_{\mathrm{x}}>0$, whereas open symbols correspond to $J_{\mathrm{x}}<0$. The colours of the symbols show the evolution during the current pulse and grey symbols show the subsequent post-pulse relaxation. \textbf{(b)} Corresponding trajectories of a reference AFM skyrmion stabilized at $\mu_0H_{\mathrm{z}}=0$ under otherwise identical conditions. Filled and open stars denote $J_{\mathrm{x}}>0$ and $J_{\mathrm{x}}<0$, respectively; grey stars show the corresponding post-pulse evolution. \textbf{(c)} Time evolution of the meron--antimeron axis orientation $\vartheta_{\mathrm{B}}$ and the in-plane background angle $\phi$ for both current polarities. The black traces show the reference background angle $\phi_{\mathrm{ref}}$ obtained in the absence of a skymeron. All three angles, $\vartheta_{\mathrm{B}}$, $\phi$, and $\phi_{\mathrm{ref}}$, are measured with respect to the positive $x$ axis, corresponding to the direction of the applied current. $\vartheta_{\mathrm{B}}$ develops a pronounced offset from $\phi$ during the pulse and continues to relax towards it after the current is switched off at $t=5~\mathrm{ns}$, revealing an internal rotational mode of the skymeron. \textbf{(d)} Time evolution of the meron--antimeron core separation $d^{\mathrm{M-A}}$, which remains nearly constant compared with the pronounced angular evolution in \textbf{c}. Insets \textbf{(i--iii)} show representative simulated magnetization configurations illustrating this reorientation. The grey shaded regions in \textbf{c,d} denote the zero-current interval after pulse termination.}
    \label{fig:4}
\end{figure}

%Our experiments reveal three characteristic stages of the skymeron dynamics: a polarity-dependent asymmetric motion during the current pulse, a short inertia-like evolution after pulse termination, and a slower return towards the initial state. The detailed deformation and long-time relaxation pathways vary between individual skymerons, reflecting the influence of the local non-flat energy landscape. To isolate the intrinsic response from these extrinsic contributions, we performed zero-temperature micromagnetic simulations of an isolated skymeron in an idealized pinning-free system.

To distinguish the intrinsic skymeron response from the influence of the local non-flat energy landscape, we performed zero-temperature micromagnetic simulations of an isolated skymeron in an idealized flat energy landscape system. Figs.~\ref{fig:4}(a,b) compare the simulated current-driven dynamics of a skymeron and a reference antiferromagnetic skyrmion under $5$-ns pulses of $|J_\mathrm{x}|=4\times10^{10}\,\mathrm{A\,m^{-2}}$ for both current polarities, with each pulse initialized from the same state (see Methods). To resolve the internal skymeron dynamics, we separately track the meron and antimeron cores of its bimeron component (Fig.~\ref{fig:4}(a)).

To demonstrate the qualitative difference, we compare to a reference AFM skyrmion that follows a symmetric longitudinal trajectory upon current reversal and exhibits negligible transverse displacement, consistent with the suppression of the skyrmion Hall effect \cite{barker2016static ,pham2024fast,bhukta2025antiferromagnetic} and shows no significant inertia. This confirms that the inertia-like response of the skymeron does not stem from AFM coupling alone, but from its internal spin texture. The skymeron instead exhibits non-rigid, polarity-dependent trajectories (Supplementary Movies~1 and 2 ). A key contribution to this non-rigid response arises from the SOT-driven rotation of the in-plane background magnetization, quantified using a reference simulation of a uniformly in-plane magnetized state in the absence of a skymeron, \(\phi_{\mathrm{ref}}\) (black trace, Fig.~\ref{fig:4}(c)). Because iDMI stabilizes the chiral configuration of the constituent textures \cite{bhukta2024homochiral}, the meron--antimeron pair reorients with the rotating background while preserving its DMI-selected chirality, resulting in a non-rigid deformation rather than a simple rigid translation. We quantify this response through the meron--antimeron axis $\vartheta_{\mathrm{B}}$ and the in-plane background angle $\phi$ (Fig.~\ref{fig:4}(c)). During the pulse, $\phi$ rapidly approaches a quasi-steady orientation, whereas $\vartheta_{\mathrm{B}}$ continues to rotate, producing a polarity-dependent angular offset. After the current is switched off, \(\vartheta_{\mathrm B}\) and \(\phi\) continue to relax towards one another, demonstrating that the internal rotational degree of freedom remains out of equilibrium after removal of the drive and provides a relaxation channel for the inertia-like post-pulse motion. By contrast, the meron--antimeron separation $d^{\mathrm{M-A}}$ remains nearly constant (Fig.~\ref{fig:4}(d)), identifying rotation rather than core separation as the principal internal degree of freedom underlying the inertia-like dynamics.

Our theoretical description shows that the \(\mathbb{Z}^{3}\) classification does not itself determine a dynamical mass; rather, in the present realization it identifies a composite skyrmion--bimeron spin texture in which the center coordinate \(\mathbf{R}\) is accompanied by internal collective coordinates, including the relative displacement of the skyrmion and bimeron components and the separation and orientation of the meron-antimeron core. Magnetic compensation suppresses the net gyrotropic force for rigid translation, but it does not fully eliminate the coupling between \(\mathbf{R}\) and these relative coordinates. For instance, in composite vortex domain walls \cite{tretiakov2007vortices, tretiakov2008dynamics}, integrating out internal degrees of freedom associated with the coordinates that are faster than the long-time center-of-mass motion, but not instantaneous on the pulse-switching timescale, generates a retarded force on \(\mathbf{R}\). At low frequencies, this appears as an effective inertial term \(M_{\mathrm{eff}}\ddot{\mathbf{R}}\), with \(M_{\mathrm{eff}}\) set by the gyro coupling squared divided by the internal restoring stiffness. For a vortex domain wall, the analogous reduction gives \(m=G^{2}/k\), while in asymmetric bimerons the different motion of the two meron constituents produces elongation and rotation \cite{tretiakov2007vortices,tretiakov2008dynamics,vorobyev2024asymmetric}.

To identify the origin of the experimentally observed post-pulse continuation, we examine the simulated response after removal of the current. The reference AFM skyrmion stops within one simulation time step, as expected from the first-order LLG equation and the resulting rigid-body Thiele description \cite{thiele1973} of a compensated antiferromagnetic skyrmion\cite{pham2024fast,bhukta2025antiferromagnetic}. However, the skymeron continues to translate and deform for approximately \(1\)–\(2\) ns, accumulating an additional displacement of approximately \(60\) nm along \(x\) and \(10\) nm along \(y\). This delayed motion is consistent with internal skyrmion-bimeron collective coordinates remaining out of equilibrium and continuing to exert a generalized force on the center coordinate as they relax. In a reduced description containing only \(\mathbf{R}\), eliminating these internal collective coordinates therefore appears as an emergent inertial response, or more generally, as a finite-memory kernel, rather than as a bare mass of a rigid quasiparticle\cite{buttner2015dynamics}. The microscopic origin of this effective memory kernel is evident in the micromagnetic energy decomposition (Extended Data Fig.~3): the anisotropy, magnetostatic, Zeeman, and interlayer-exchange contributions undergo a substantial redistribution during the current pulse and relax towards their equilibrium values only on a finite timescale after the drive is removed.

The initial $\sim1$--$2~\mathrm{ns}$ post-pulse motion observed experimentally (region (i) in Fig.~\ref{fig:3}(c)) is consistent with the simulations, supporting its origin in the finite-time relaxation of the internally deformed skymeron. We note that the simulations are performed at zero temperature, which limits a fully quantitative comparison despite the use of experimentally determined parameters (see Methods). In addition, the effective damping extracted from the experiment\cite{dohi2024observation} contains multiple contributions, and the relative roles of the Oersted field and field-like torque cannot be disentangled. Within these limitations, the simulations reproduce the key experimental features and identify internal relaxation as the microscopic origin of the inertia-like response, which is unique to the skymeron and qualitatively different from conventional spin textures such as skyrmions.

\section*{Conclusion}

In conclusion, we realize and directly image the skymeron, a three-dimensional composite spin texture consisting of an out-of-plane skyrmion in one layer coupled to an in-plane bimeron in the other. This state is enabled by a degree of freedom in materials design that is specific to synthetic antiferromagnets: the independent engineering of the effective anisotropies of the two antiferromagnetically exchange-coupled ferromagnetic layers. In an out-of-plane magnetic field, the anisotropy-energy imbalance between the two layers drives an asymmetric spin-flop transition, thereby transforming an antiferromagnetic skyrmion into a skymeron. For the topology, the transition from the polar–polar \(\mathbb Z^{2}\) sector to the polar–planar \(\mathbb Z^{3}\) sector has both a direct experimental signature and a dynamical consequence. Current pulses excite internal collective coordinates of the skymeron, producing polarity-dependent deformation and non-rigid motion. Their finite-time relaxation gives rise to the inertia-like post-pulse response, whereas the slower reverse motion is governed predominantly by the local non-flat energy landscape. The skymeron therefore provides internal dynamical degrees of freedom beyond the rigid-body description of compensated antiferromagnetic skyrmions. Thus, topology identifies and constrains the composite constituents that support the internal modes, while their energetic stiffness, damping, and the local pinning landscape determine the effective inertia, relaxation time, and trajectory. Most importantly, our work establishes anisotropy engineering in coupled magnetic order parameters as a route for designing unique composite spin textures. This concept is not restricted to sputtered synthetic antiferromagnets: independent tuning of the anisotropy \cite{fallon2026unconventional}, intralayer DMI \cite{bhukta2022degenerate}, and Heisenberg exchange of the constituent layers may enable further unconventional textures with distinct dynamical responses. Unconventional spin textures with complex dynamics may thus be realized in compensated ferrimagnets and van der Waals magnetic heterostructures, where interlayer exchange and Moiré potentials provide additional control parameters.

\section*{Acknowledgments}

We thank the Helmholtz-Zentrum Berlin für Materialien und Energie for the allocation of synchrotron radiation beamtime. M.K. acknowledges the Lawrence Berkeley National Laboratory for the hospitality.

\section*{Funding}
This research used resources of the Advanced Light Source, which is a DOE Office of Science User Facility under contract no. DE-AC02-05CH1123. This work has received funding from the European Union’s Horizon 2020 research and innovation program under the Marie Skłodowska-Curie Grant Agreement No. 860060 “Magnetism and the effect of Electric Field” (MagnEFi), as well as from Synergy Grant No. 856538, project “3D-MAGiC,” the Horizon Europe Project No. 101070290 (NIMFEIA), and Horizon Europe Programme Horizon 1.2 under the Marie Skłodowska-Curie Actions (MSCA), Grant agreement No.101119608 (TOPOCOM). It has also been supported by the Deutsche Forschungsgemeinschaft (DFG, German Research Foundation) – SPP 2137 Skyrmionics (project 462597720), TRR 173 – 268565370 (projects A01, A12 and B02), and the Dynamics and Topology Center TopDyn funded by the State of Rhineland Palatinate. This work was also partly supported by the Norwegian Research Council through its Center of Excellence, Project Number 262633, “QuSpin”. F.N.R. and O.E. acknowledge support from the Swedish Research Council (Grant No. 2023-04899). O.A.T. acknowledges support from the Australian Research Council (Grant No. DP240101062) and the NCMAS grant. This work was partly supported by JSPS Kakenhi (Grant Nos. 23K13655, 24H00039, 24H00409, 25K01645) as well as the Center for Science and Innovation in Spintronics at Tohoku University (Cooperative Research Project). This work was partly supported by the National Research Council of Science \& Technology (NST) grant by the Korean government, MSIT (Grant No. GTL24041-000).

%\section*{Author Contributions}

%M.B., R.F., and M.K. conceived and initiated the project.  N.S.K. and F.N.R. led the theoretical analysis and developed the topological classification with input from O.E. T.D., D.M.T., and M.A.S. deposited the multilayer stack. M.B. and D.M.T. characterized the samples by SQUID magnetometer and MFM imaging, fabricated the devices by electron-beam lithography, and fabricated the electrical contacts. M.B. supervised the STXM experiments and analysed the STXM data with support from S.W., M.W., and H.O. M.B., D.M.T, F.K., T.D., R.T., E.M., B.F., and S.K. participated in the beamtime experiments. T. Denneulin, J.V.V., and R.E.D. conducted and analysed the HAADF-STEM and EDX measurements. K.L. and M.B. performed the micromagnetic simulations with the support of O.T. M.B. drafted the manuscript with input from N.S.K., F.N.R., R.F., and M.K. All authors discussed the results and commented on the manuscript.

\section*{Competing Interests}
The authors declare no conflict of interest.

\section*{Methods}

\subsection*{Topological analysis}

\begin{table}[h]
\caption{
\label{table_glossary}
Table of symbols and terms.
}
\begin{tabular}{p{14mm} p{60mm} }
\hline
\multicolumn{1}{l}{Symbol} & \multicolumn{1}{c}{Meaning}   \\
\hline\hline
$\cong$ & homeomorphic \\
$\simeq$ & homotopy equivalent \\
$\times$ & Cartesian product (spaces) or direct product (groups)  \\
$\rtimes$ & semidirect product of groups  \\
$\backslash$ & set-theoretic difference  \\
$\subseteq$ & subset or equal to \\
$\rightarrow$ & continuous map (spaces) or homomorphism (groups)  \\
$\mapsto$ & maps to   \\
$\hookrightarrow$ & inclusion map  \\
$\partial X$ & boundary of a space~$X$  \\
$\pi_n(X)$ & $n$-th homotopy group of~$X$  \\
$\pi_n(X, Y)$ & $n$-th relative homotopy group of $X$ modulo~$Y$  \\
$\mathrm{Aut}(X)$ & automorphism group of~$X$  \\
$\mathrm{cap}$ & spherical cap  \\
$\mathrm{Conf}_n(X)$ & $n$-th configuration space of~$X$  \\
$\mathrm{GL}_n(\mathbb{Z})$ & group of ${n\times n}$ integer matrices that have integer inverses  \\
$\mathrm{segment}$ & spherical segment surface  \\
$\mathbb{S}^n$ & $n$-dimensional sphere  \\
$\mathbb{Z}$ & group of integers under addition  \\
$\mathbb{Z}^n$ & free Abelian group of rank~$n$  \\
\hline
\end{tabular}
\end{table}

Topological invariants were defined using methods of algebraic topology~\cite{Hatcher}. 
The classification of localized spin textures is formulated in terms of relative homotopy groups of pairs of spaces.
In a two-sublattice SyAFM, the order parameter consists of two unit magnetization vectors, and therefore takes values in the product space
$X = \mathbb{S}^2 \times \mathbb{S}^2$.
A localized magnetic texture occupies a two-dimensional domain that can be continuously transformed into (in other words, it is homeomorphic to) a disk $\mathbb{D}^2$ with boundary $\partial\mathbb{D}^2$. 
The topological classification~\cite{Rybakov2025} is therefore formulated in terms of maps of pairs
$f : (\mathbb{D}^2, \partial\mathbb{D}^2) \to (X, Y),$
where $Y \subseteq X$ denotes the subset of order-parameter values realized at the boundary~$\partial\mathbb{D}^2$. 
The physical meaning of the subset $Y \subseteq X$ is that it represents only the energetically preferred spin configurations, taken into consideration with finite angular extent to account for moderate deviations from the ideal homogeneous ground state.
The relevant topological invariant is within the relative homotopy group~$\pi_2(X, Y)$. 
To determine it, we use the long exact sequence of homotopy groups for the pair~$X \supseteq Y$:

\begin{numcases}{}
\dots
\rightarrow
\pi_2(Y)
\xrightarrow{i_2}
\pi_2(X)
\xrightarrow{j_2}
\pi_2(X, Y)
\rightarrow
\pi_1(Y)
\rightarrow
\dots \label{sequence_long} \\
\text{im}\,j_2  \subseteq Z\left( \pi_2(X, Y) \right), \label{sequence_j2}
\end{numcases} 
where constraint~(\ref{sequence_j2}) is of geometric nature~\cite{Hilton1953} and means that the image of homomorphism $j_2$ lies in the center of group $\pi_2(X, Y)$. 

Sequence~(\ref{sequence_long}) with extra condition~(\ref{sequence_j2}) can be interpreted as a system of equations for groups. 
The arrows denote group homomorphisms, while exactness means that the image of every homomorphism equals the kernel of the next.
Since
$\pi_2(X)=\pi_2(\mathbb{S}^2 \times \mathbb{S}^2)=
\mathbb{Z} \times \mathbb{Z}$,
the objective group $\pi_2(X,Y)$ is entirely determined by the choice of boundary subspace~$Y$, as it defines groups $\pi_{1,2}(Y)$ in Eq.~(\ref{sequence_long}).

If the sublattices are independent and do not interact, then the group is a direct product. 
In particular, for the case where two sublattices are easy-plane and independent, we get 
\begin{align}
\label{ZZZZ}
&\pi_2(
\mathbb{S}^2 
\times 
\mathbb{S}^2, 
\mathbb{S}^2\setminus\{P_1,P_2\} 
\times 
\mathbb{S}^2\setminus\{P_1,P_2\}
) = \nonumber \\
&\pi_2(
\mathbb{S}^2
,
\mathbb{S}^2\setminus\{P_1,P_2\}
)\times
\pi_2(
\mathbb{S}^2
,
\mathbb{S}^2\setminus\{P_1,P_2\}
) = \nonumber \\
&\pi_2(
\mathbb{S}^2
,
\mathbb{S}^1
)\times
\pi_2(
\mathbb{S}^2
,
\mathbb{S}^1
) = \nonumber \\
&(\mathbb{Z}\times\mathbb{Z})
\times
(\mathbb{Z}\times\mathbb{Z})
\equiv
\mathbb{Z}^4.
\end{align}
This auxiliary relation is useful for deriving several of the following formulas.
%The derivation of group~(\ref{ZZZZ}) follows from the results of Ref.~\cite{Rybakov2025} %Due to the following citation we don't need this sentence anymore
According to the terminology of Ref.~\cite{Rybakov2025} corresponding topological charge is a quadruplet of so-called \textit{top} and \textit{bottom} indices: $(q^\text{(A)}_\text{t}, q^\text{(A)}_\text{b}, q^\text{(B)}_\text{t}, q^\text{(B)}_\text{b})$. 
The integer invariants $q^{(i)}_{j}$ were computed following the procedure described in Ref.~\cite{Rybakov2025}, by means of an auxiliary dumbbell map applied to magnetization in both sublattices~A and~B.
%, respectively. 
%
Some additional details on the implementation of the dumbbell map and calculation of $q_\text{t}$ and $q_\text{b}$ can be found in Refs.~\cite{Bimerons2025, chen2026_hopfion_rings}.
Care was taken to maintain consistent orientation conventions in order to avoid ambiguities associated with polarity choices.

\noindent\textbf{Case I: Polar-polar antiferromagnetic background.}
For strong perpendicular anisotropy and low magnetic field, it is energetically preferable to be in a near-polarized state in both sublattices, see Fig.~\ref{fig:1}(d). 
Therefore, we imply that~$Y$ is the product of spherical caps, but this space in turn is homotopic to the point: $Y\cong\mathrm{cap}\times\mathrm{cap}\simeq\mathrm{point}\times\mathrm{point}\simeq\mathrm{point}$. 
Accordingly, $\pi_2(Y)=0$ and $\pi_1(Y)=0$, and the long exact sequence~(\ref{sequence_long}) reduces to the short exact sequence
\begin{align}
0
\rightarrow
\mathbb{Z}\times\mathbb{Z}
\xrightarrow{j_2}
\pi_2(X,Y)
\rightarrow 0.
\end{align}
This has a unique solution
\begin{align}
\pi_2(\mathbb{S}^2 \times \mathbb{S}^2, \mathrm{cap}\times\mathrm{cap})=\mathbb{Z} \times \mathbb{Z}.   
\label{Z2}
\end{align}
The corresponding integer pair
$(Q^{\mathrm{(A)}}, Q^{\mathrm{(B)}})$
represents the skyrmion numbers of the two sublattices, A~and~B. 
These configurations correspond to antiferromagnetic skyrmions, as shown schematically in Fig.~\ref{fig:1}(a).
The inclusions 
\begin{align}
\mathbb{S}^1
\times 
\mathbb{S}^1
\hookrightarrow
\mathrm{cap}\times\mathrm{cap}
\end{align}
induce group homomorphisms from~(\ref{ZZZZ}) to~(\ref{Z2}). 
In particular, if sublattice~A is polarized upwards, and using the formulas for top-bottom indices~\cite{Rybakov2025}:
\begin{align}
(q^\text{(A)}_\text{t}, q^\text{(A)}_\text{b}, q^\text{(B)}_\text{t}, q^\text{(B)}_\text{b}) \mapsto
(q^\text{(A)}_\text{b}, q^\text{(B)}_\text{t}) = 
(Q^{\mathrm{(A)}}, Q^{\mathrm{(B)}}). 
\end{align}
For instance, for the skyrmion depicted in Fig.~\ref{fig:1}(a), we obtain $(Q^{\mathrm{(A)}}, Q^{\mathrm{(B)}})=(-1,-1)$.

\noindent\textbf{Case II: Polar-planar background.}
Under an increasing magnetic field, one sublattice may remain nearly polarized while the other acquires easy-plane (planar) behavior, see Fig.~\ref{fig:1}(f). 
In this case 
$Y\cong\mathrm{cap}\times\mathrm{segment}\simeq\mathrm{point}\times\mathbb{S}^1\simeq\mathbb{S}^1$.
Substituting this into (\ref{sequence_long}) yields the short exact sequence
\begin{align}
\label{sequence_short_b}
0
\rightarrow
\mathbb{Z}\times\mathbb{Z}
\xrightarrow{j_2}
\pi_2(X,Y)
\rightarrow
\mathbb{Z}
\rightarrow
0.
\end{align}
The general solution of this short split exact sequence is the following group:~$(\mathbb{Z}\times\mathbb{Z}) \rtimes \mathbb{Z}$.
However, 
%geometric considerations imply that the image of $j_2$ lies in the center of $\pi_2(X, Y)$, so 
constraint~(\ref{sequence_j2}) consequently implies that 
only the trivial action of~$\mathbb{Z}$ is possible in this case.
%compatible with the geometric construction inducing sequence~(\ref{sequence_short_b}).
%
%Consequently
As a result, the semidirect product reduces to a direct product:
\begin{align}
\label{Z3}
\pi_2(\mathbb{S}^2\times\mathbb{S}^2,
\mathrm{cap}\times\mathrm{segment})
=
(\mathbb{Z}\times\mathbb{Z})\times\mathbb{Z}
\equiv \mathbb{Z}^3.
\end{align}
Thus, spin textures in this regime are characterized by a triplet of integer invariants. 
The inclusions 
\begin{align}
\mathbb{S}^1
\times 
\mathbb{S}^1
\hookrightarrow
\mathrm{cap}\times\mathrm{segment}
\end{align}
induce group homomorphisms from~(\ref{ZZZZ}) to~(\ref{Z3}). 
In particular, if sublattice~A is polarized upwards:
\begin{align}
(q^\text{(A)}_\text{t}, q^\text{(A)}_\text{b}, q^\text{(B)}_\text{t}, q^\text{(B)}_\text{b}) \mapsto
(q^\text{(A)}_\text{b}, q^\text{(B)}_\text{t}, q^\text{(B)}_\text{b}).
\end{align}
Thus, the skymeron sketched in Fig.~\ref{fig:1}(c) has a topological charge $(q^\text{(A)}_\text{b}, q^\text{(B)}_\text{t}, q^\text{(B)}_\text{b}) = (-1, -1, 1)$.
Moreover, we need to note that the above triple of integers may be expressed in (infinitely) many equivalent ways.
% , but
Namely, in accordance with automorphisms of the free abelian group~(\ref{Z3}), 
$\mathrm{Aut}(\mathbb{Z}^3) = \mathrm{GL}_3(\mathbb{Z})$.
In this regard, one of the practically useful automorphism is
$
M =
\begin{pmatrix}
1 & -1 & 0 \\
0 & 1 & 0 \\
0 & 0 & 1
\end{pmatrix}
\in \mathrm{GL}_3(\mathbb{Z}),
$
which yields
\begin{align}
M \cdot 
(q^\text{(A)}_\text{b},\,q^\text{(B)}_\text{t},\,q^\text{(B)}_\text{b})^\text{T} 
= 
(\underbrace{q^\text{(A)}_\text{b} - q^\text{(B)}_\text{t}}_\text{sing. index},\,q^\text{(B)}_\text{t},\,q^\text{(B)}_\text{b})^\text{T} 
\end{align}
The first component in this transformed triple explicitly represents the index of singularity, 
which can be interpreted as the antiferromagnetic analogue of a point defect in 
a ferromagnetically coupled bilayer~\cite{Ghosh2026}. 
For the skymeron shown in Fig.~\ref{fig:1}(c), the index of singularity equals $-1 - (-1) = 0$, and the texture is therefore non-singular.

\noindent\textbf{Transition between cases~I~and~II.}
The inclusions
\begin{align}
%$\mathrm{pt}\times\mathbb{S}^1 \hookrightarrow \mathrm{pt}\times\mathrm{pt}$
\mathrm{cap}\times\mathrm{segment}
\hookrightarrow
\mathrm{cap}\times\mathrm{cap}
\end{align}
induce a homomorphism between the corresponding relative homotopy groups:
\begin{align}
\label{homom}
\mathbb{Z}^3
\rightarrow
\mathbb{Z}^2,
\quad
(q^\text{(A)}_\text{b}, q^\text{(B)}_\text{t}, q^\text{(B)}_\text{b})
\mapsto
(q^\text{(A)}_\text{b}, q^\text{(B)}_\text{t})= 
(Q^{\mathrm{(A)}}, Q^{\mathrm{(B)}}).
\end{align}
This mapping reflects the physical transition from the polar-planar boundary regime to the antiferromagnetic polar-polar background.
Topologically, this corresponds to the continuous transformation of a skymeron into an antiferromagnetic skyrmion.

\subsection*{Thin-film deposition and device fabrication}

The SyAFM multilayer comprises  1.1-\(\mathrm{nm}\)-thick \(\mathrm{Co}_{80}\mathrm{B}_{20}\) (CB) layers forming the ferromagnetic sublattice B and \(\mathrm{Co}_{60}\mathrm{Fe}_{20}\mathrm{B}_{20}\)/ \(\mathrm{Fe}_{60}\mathrm{Co}_{20}\mathrm{B}_{20}\) (CFB/FCB) bilayers forming sublattice A. The Pt and Ir interfaces break inversion symmetry and induce a strong interfacial DMI, stabilizing homochiral spin textures within each FM sublattice while enabling control of the interlayer exchange coupling. The CFB and FCB layers are \(0.35\,\mathrm{nm}\) and \(0.375\,\mathrm{nm}\) thick, respectively. This compositional asymmetry enables element-selective XMCD imaging: the Fe \(L_3\) edge selectively probes the Fe-containing sublattice A, whereas the Co \(L_3\) contrast is dominated by the Co-rich sublattice B. From superconducting quantum interference device
(SQUID)-magnetometry, we obtain a magnetic compensation \(m_{\mathrm{c}} = 1-\left|M_{\mathrm{s,A}}t_{\mathrm{A}}-M_{\mathrm{s,B}}t_{\mathrm{B}}\right|/(M_{\mathrm{s,A}}t_{\mathrm{A}}+M_{\mathrm{s,B}}t_{\mathrm{B}})\) of approximately \(91\%\). Here \(t_{\mathrm{A}}\) and \(t_{\mathrm{B}}\) denote the total ferromagnetic thicknesses of sublattices A and B, respectively, and \(M_{\mathrm{s,A}}\) and \(M_{\mathrm{s,B}}\) are their corresponding saturation magnetizations. The saturation magnetizations of the individual sublattices are \(M_{\mathrm{s,A}}=(1500\pm120)\,\mathrm{kA\,m^{-1}}\) and \(M_{\mathrm{s,B}}=(820\pm120)\,\mathrm{kA\,m^{-1}}\). Supplementary Notes 1 and 2 provide the details for the multilayer stack optimization procedure. The multilayer stacks investigated in this study were deposited using a Singulus Rotaris magnetron-sputtering system at a base pressure of \( 3 \times 10^{-8} \, \text{mbar} \) at room temperature, using Ar as the sputtering gas. Cross-sectional transmission electron microscopy confirms the intended multilayer architecture, with continuous layers and well-defined interfaces across the stack (Extended Data Fig.~1). Static measurements were performed on the stack described above, whereas the time-resolved current-driven measurements were performed at an out-of-plane magnetic field of $\mu_0H_{\mathrm{z}}=120$ $ \mathrm{mT}$ on a second stack in which the Pt layer thickness was changed to (1.25 nm); all other layer thicknesses were unchanged. 

%To minimize pinning effects, low deposition power and low argon flow were used,\ resulting in smooth and homogeneous individual layers verified using X-ray reflectivity (XRR) and cross-sectional transmission electron microscopy (TEM) measurements (See Supplementary Note 2). 

The SyAFM multilayers were deposited onto a 100-nm-thick silicon nitride membrane window. Track structures were defined using electron-beam lithography on a bilayer resist, followed by lift-off. In a second lithography step, sputter-deposited Cr (5 nm)/Au (200 nm) contact pads were patterned at both ends of the tracks to enable injection of electrical current pulses. A 200-nm-thick Al layer was deposited on the back side of the membrane to improve heat dissipation during current pulsing. The high structural quality of our SyAFM stack, confirmed by cross-sectional TEM of the multilayer stack (Extended Data Fig.~1), indeed enabled the skymeron's unconventional dynamics to be resolved with full reproducibility across billions of excitation cycles.

\subsection*{Static XMCD-STXM Imaging and line-profile analysis}
The X-ray magnetic circular dichroism (XMCD) imaging experiments were performed at the MAXYMUS~\cite{weigand2022timemaxyne} endstation at the BESSY~II electron storage ring, operated by the Helmholtz-Zentrum Berlin für Materialien und Energie~\cite{follath2010x}, and at the scanning transmission X-ray microscope of the Advanced Light Source (ALS), Lawrence Berkeley National Laboratory. XMCD imaging exploits the helicity-dependent absorption of circularly polarized X-rays, which is sensitive to the projection of the magnetization along the X-ray propagation direction. This technique provides element-specific contrast by tuning the incident X-ray energy to the absorption edges of transition metals, allowing direct imaging of the local magnetic configuration. In this study, time-resolved scanning transmission X-ray microscopy (STXM) was employed to capture the real-space magnetization profile. X-ray absorption spectra were recorded prior to imaging, confirming the resonance conditions for Fe and Co at photon energies of nominally 708.1 eV and 778.6 eV, respectively. The unambiguous contrast inversion between the sublattices at zero field verified the strong antiferromagnetic coupling, as observed in static and dynamic imaging sequences. The stability of this coupling was further confirmed by maintaining the characteristic XMCD contrast throughout the entire range of applied current densities.

Line profiles in Fig. \ref{fig:2} were extracted along the directions W1 and W2 and averaged over 25 pixels perpendicular to the profile direction to improve the signal-to-noise ratio at 25 nm spatial resolution. Time-resolved pump-probe measurements were performed to capture the nanosecond-scale dynamics of the skymerons. From the spatial extent of the XMCD contrast, we deduce for our system a characteristic skymeron radius of $(150 \pm 25)\,\mathrm{nm}$. The line profile along $W_2$ further reveals the directional character of the in-plane background in layer~B. Whereas the stripe probed by $W_1$ corresponds to the standard $360^\circ$ rotation associated with a bimeron, the profile along $W_2$ indicates an effective rotation of approximately $540^\circ$ across the stripe. The in-plane background magnetization on the two sides of this domain wall is therefore antialigned. Consistently, the skymerons located on opposite sides of this wall (labeled S1 and S2 in Fig.~\ref{fig:1}(k)) exhibit inverted black--white and white--black core contrasts in layer~B. This inversion reflects the reversal of the surrounding in-plane background magnetization, while the chirality of the two textures remains identical owing to the homochiral nature of the bimerons in the system~\cite{bhukta2024homochiral} induced by interfacial Dzyaloshinskii–Moriya interaction (DMI) \cite{dzyaloshinsky1958thermodynamic, moriya1960anisotropic}. We note that STXM provides a thickness-averaged magnetic contrast across the entire multilayer stack. Consequently, the Co signal contains contributions from both layers A and B. However, owing to the substantially higher Co content in layer B (See Supplementary Note 2), the observed contrast is dominated by layer B.

The nucleation of skymerons discussed above is governed by the local energy landscape, where defect-induced energy barriers stabilize the textures at specific positions. To obtain isolated skymerons with improved spatial control, we employ current-induced nucleation. Current-driven creation of topological textures has previously been demonstrated for both ferromagnetic \cite{lemesh2018current,woo2016observation} and antiferromagnetic skyrmions \cite{dohi2024observation,bhukta2025antiferromagnetic}, providing a controlled pathway for generating individual spin textures (see Supplementary Note~3).

\subsection*{Time-resolved STXM Imaging}

%Short X-ray probe pulses (70–100 ps) from the storage ring are synchronized with the electrical excitation, allowing the magnetic configuration to be sampled at well-defined time points within the excitation cycle. Because the measured magnetic contrast represents an average over billions of identical pump–probe cycles, only reproducible skymeron trajectories contribute to the observed signal. Non-deterministic motion averages out and remains invisible in the experiment. In addition, the transmission-based imaging geometry integrates the magnetic signal over all repeated multilayers. 

The time-resolved pump-probe experiment utilized a 2001-channel time-resolved detection scheme, enabling precise reconstruction of skymeron motion over the entire excitation cycle. A bipolar current pulse sequence was applied, with a 1 $\mu$s delay between successive positive and negative pulses, ensuring reproducible skymeron trajectories without cumulative heating effects (See Supplementary Note 6). The measurements were conducted nominally at room temperature and in a 10-mbar helium atmosphere to allow for some thermal convection, mitigating heating-related artifacts. 

For the analysis of the time-resolved dynamics shown in Fig.~\ref{fig:3}(a), the skyrmion core was segmented from each XMCD image and its closed boundary was extracted. At each longitudinal position, the transverse coordinates of the upper and lower boundary branches were averaged to obtain the one-dimensional core centreline shown in Fig.~\ref{fig:3}(a). The resulting centreline was spatially smoothed to suppress pixel-scale variations. Successive centrelines were vertically offset and stacked to visualize their evolution throughout the excitation sequence, with colour encoding time. The green ($0$--$10$ ns) and blue ($1000$--$1010$ ns) shaded regions mark the positive and negative current pulses, respectively, whereas the subsequent traces show the corresponding post-pulse evolution. The two skymerons, denoted by a circle and a downward-pointing triangle, are tracked throughout the pump-probe cycle.

%For the time-resolved movie analysis shown in Fig.~\ref{fig:3}(a), the skyrmion constituents of the skymeron were segmented from each XMCD image and its closed boundary was extracted. At each longitudinal position, the transverse coordinates of the upper and lower boundary branches were averaged to obtain a one-dimensional core centreline. The resulting centreline was spatially smoothed to suppress pixel-scale variations. Successive centrelines were vertically offset and stacked to visualize their temporal evolution throughout the bipolar excitation cycle, with colour encoding the corresponding time.

\subsection*{Micromagnetic simulations}
\begin{table}[h]
\centering
\begin{tabular}{lcccc}
\hline
Parameter & $\mathrm{FM}_\mathrm{A}$ & $\mathrm{FM}_\mathrm{B}$ & $\mathrm{FM}_\mathrm{A'}$ & $\mathrm{FM}_\mathrm{B'}$ \\
\hline\hline
Saturation magnetization $M_{\mathrm{s}}$ (\si{\kilo\ampere\per\meter}) 
& 840 & 840 & 840 & 840 \\

Exchange stiffness $A$ (\si{\pico\joule\per\meter}) 
& 10 & 10 & 10 & 10 \\

Uniaxial magnetocrystalline anisotropy along $\vec{e}_z$, $K$ (\si{\kilo\joule\per\cubic\meter}) 
& 510 & 400 & 520 & 400 \\

Interfacial DMI strength $D$ (\si{\milli\joule\per\square\meter}) 
& 0.47 & 0.47 & 0.54 & 0.54 \\

Interlayer exchange coupling $J_{\mathrm{IEC}}$ (\si{\milli\joule\per\square\meter}) 
& -0.12 & -0.12 & -0.12 & -0.12 \\
\hline
\end{tabular}
\caption{Micromagnetic parameters used in the simulations. Note that for $\mathrm{FM}_\mathrm{B}$ and $\mathrm{FM}_{\mathrm{B'}}$ the effective anisotropy $K_{\mathrm{eff}}=K-\mu_0 M_{\mathrm{s}}^2/2$ is negative, whereas for $\mathrm{FM}_\mathrm{A}$ and $\mathrm{FM}_\mathrm{A'}$ it is positive.}
\label{tab:micromagparameters1}
\end{table}

The micromagnetic simulations were performed using a modified version of \texttt{MuMax3}~\cite{vansteenkiste2014design} that allows to specify the thickness of each ferromagnetic layer as well as the thickness of each non-magnetic spacer layer (i.e., the distance between adjacent ferromagnetic layers) explicitly, including the latter in the magnetostatic calculations without storing non-magnetic regions in memory as equidistant cells with $M_{\mathrm{s}}=0$. The coordinates $x$ and $y$ denote the in–plane directions of the multilayer system, while $z$ represents the direction perpendicular to the layers. In the $x$ and $y$ directions the system was discretized into $N_\mathrm{x}=N_\mathrm{y}=256$ cells with a cell size of $\Delta_\mathrm{x}=\Delta_\mathrm{y}=\SI{4} 
 {\nano\meter}$. This corresponds to a simulated area of $\SI{1024}{\nano\meter} \times \SI{1024}{\nano\meter}$. For the magnetostatic kernel, $P_\mathrm{x}=P_\mathrm{y}=30$ periodic images were included on each side of the simulation box in the $x$ and $y$ directions. For the skymeron, a multilayer stack of the form \(\left[(\mathrm{FM}_{\mathrm{A}},\SI{0.9}{\nano\meter}),(\mathrm{NM},\SI{1.6}{\nano\meter}),(\mathrm{FM}_{\mathrm{B}},\SI{0.9}{\nano\meter}),(\mathrm{NM},\SI{1.6}{\nano\meter})\right]\times 25\) was simulated, where FM denotes a ferromagnetic layer and NM a non-magnetic spacer layer. The index specifies the magnetic material used, with the corresponding micromagnetic parameters listed in \autoref{tab:micromagparameters1}. For the skyrmion simulations, the same multilayer structure was used, with $\mathrm{FM}_{\mathrm{A}'}$ replacing $\mathrm{FM}_{\mathrm{A}}$ and $\mathrm{FM}_{\mathrm{B}'}$ replacing $\mathrm{FM}_{\mathrm{B}}$. The system was simulated, including dipolar interactions, Heisenberg exchange, uniaxial magnetocrystalline anisotropy along the $\vec{e}_\mathrm{z}$ axis, interfacial isotropic DMI, and interlayer exchange coupling. The corresponding micromagnetic parameters are summarized in \autoref{tab:micromagparameters1}. Static skymeron simulations were performed at the respective magnetic fields $\vec{B}=B_\mathrm{z} \vec{e}_\mathrm{z}$ indicated in the figures. The skymeron dynamics were simulated at an applied magnetic field of $B_\mathrm{z}=\SI{200}{\milli\tesla}$, whereas the skyrmion dynamics were simulated at $B_\mathrm{z}=\SI{0}{\tesla}$. All simulations were carried out at $T=\SI{0}{\kelvin}$. 

%Also, since our micromagnetic calculations are done at zero temperature, we cannot expect a full quantitative agreement. Therefore, our calculations are used to provide the key qualitative explanations of the major features in the experimental observation.

The magnetic configurations $\vec{m}(\vec{r})$ were initialized using suitable ansatz states and subsequently relaxed by energy minimization using the method described in Ref.~\cite{exl}. For the dynamical simulations, spin–orbit torques were included using the Landau–Lifshitz–Gilbert–Slonczewski equation~\cite{SLONCZEWSKI1996L1,Buettner2017,vansteenkiste2014design}, with a Gilbert damping parameter $\alpha=0.1$, consistent with experimental values. %The spin polarization direction is given by the unit vector $\vec{p}=\vec{j}\times\vec{e}_\mathrm{z}$. 
The damping–like spin–orbit torque coefficient is given by $a_j=\frac{\hslash \theta_{\mathrm{SH}} |\vec{j}|}{2 M_{\mathrm{s}} e t}$, where $t=\SI{0.9}{\nano\meter}$ is the thickness of the magnetic layer, $|\vec{j}|$ is the applied current density, $e$ the elementary charge, $\hslash $ the reduced Planck constant, and $\theta_{\mathrm{SH}}=0.1$ the spin Hall angle. The field–like torque coefficient was taken as $b_j=-0.15\,a_j$. The applied current density was $\vec{j}=\pm\SI{4e10}{\ampere\per\square\meter}\,\vec{e}_\mathrm{x}$ for a duration of \SI{5}{\nano\second}. Afterwards, the current was switched off ($\vec{j}=0$), and the system was simulated for an additional \SI{5}{\nano\second}. Subsequently, the resulting simulation data were used for analysis. As a consistency check, additional simulations were performed using the standard \texttt{MuMax3} version~\cite{vansteenkiste2014design} together with the effective ferromagnetic medium model~\cite{woo2016observation}. In this approach, each ferromagnetic layer is simulated explicitly but rescaled to an effective thickness of \SI{2.5}{\nano\meter} (corresponding to 50 cells of \SI{2.5}{\nano\meter} in the $z$ direction), such that no explicit non-magnetic spacers are present between the ferromagnetic layers. No significant differences in the results were observed.

The reference AFM skyrmion in Fig.\ref{fig:4} is obtained by setting $\mu_0 H_\mathrm{z} = 0$ mT, where the two sublattices retain the conventional polar--polar configuration. By contrast, the skymeron was stabilized at $\mu_0H_{\mathrm{z}}=200$ mT, with the in-plane background of its bimeron component initially oriented along $+y$, perpendicular to the current direction, as in the experimental configuration.  In both cases, a $5$-$\mathrm{ns}$ current pulse of $J_\mathrm{x} = 4 \times 10^{10}\,\mathrm{A\,m^{-2}}$ is applied, followed by relaxation after the current is switched off. The response is evaluated for both current polarities, $J_{\mathrm{x}}>0$ and $J_{\mathrm{x}}<0$, with the skymeron or AFM skyrmion reset to the same initial state at (0,0) nm before each pulse. The background magnetization, $\phi_{\mathrm{ref}}$ (black trace in Fig.~\ref{fig:4}(c)), can be approximated by assuming one monodomain per layer, with two layers and one monodomain in each layer, using the LLG--Slonczewski equation with SOT and the effective fields of interlayer exchange and effective anisotropy. No significant force from the skymeron is exerted on the background.

%To determine the optimal parameters $(a^*, b^*)$, we consider different interaction potentials parameterized by $a$ and $b$, for which we numerically integrate the Thiele equation and obtain the corresponding trajectories for the L and R classes of skyrmions. The estimated interaction parameters $(a^*, b^*)$ are those for which $L(a,b)$ reaches its minimum, and the uncertainty is that which results from the loss landscape around the minimum $(a^*, b^*)$ (see supplementary section 1). This approach constitutes a direct determination of the interaction potential from the real-time relaxation of nanosecond scattering trajectories, without relying on indirect methods of determination. The best-fit parameters are $a^* = (38 \pm 5)~\mathrm{nm}$ and $b^* = (236 \pm 9)~\mathrm{nm}$. 
\section*{Data availability}
The data supporting the findings of this study are available within the article and Supplementary Information. Source data are provided on Zenodo.

\section*{Code availability}
The computer codes used for data analysis are available upon reasonable request from the corresponding author.

\newpage
\bibstyle{nature}
\bibliography{getwriting}

\begin{figure}[h!]
    \centering
    \includegraphics[width = 17cm]{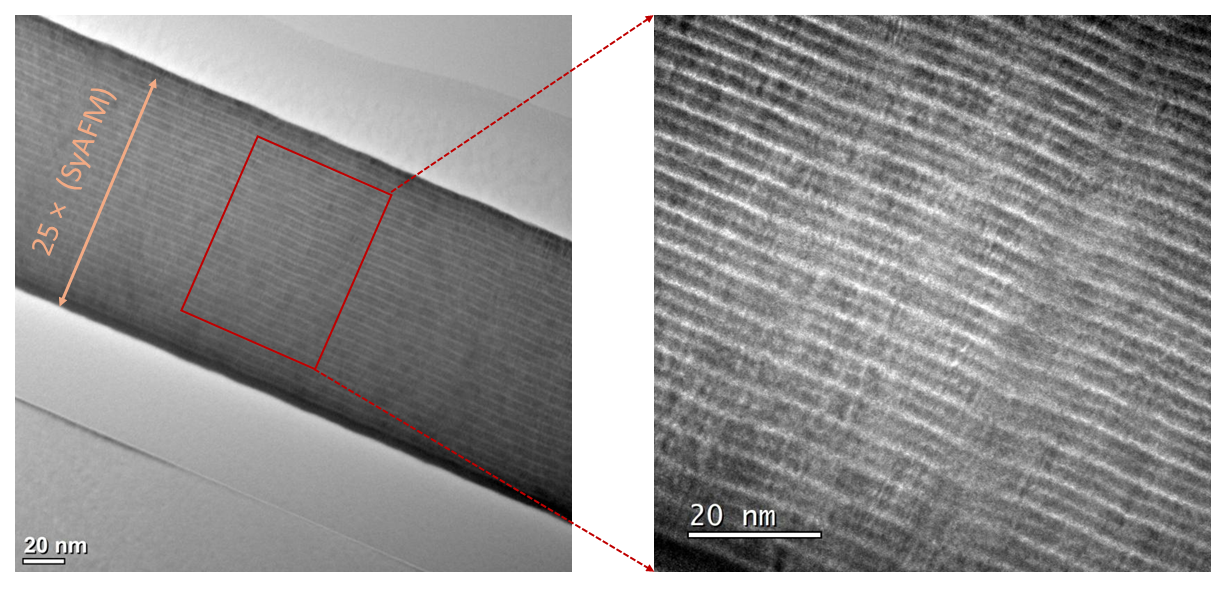}
      \captionsetup{width=0.99\textwidth}
        \caption*{\textbf{Extended Data Figure 1. Structural analysis of the SyAFM stack using cross-sectional TEM}. \textbf{(a)} Cross-sectional TEM image of the multilayer showing 25 repetitions of the SyAFM unit. The magnified view indicates the well-defined multilayer periodicity and uniform layer thickness. 
        }
        \label{figextended1}
\end{figure}

\begin{figure}[h!]
    \centering
    \includegraphics[width = 17cm]{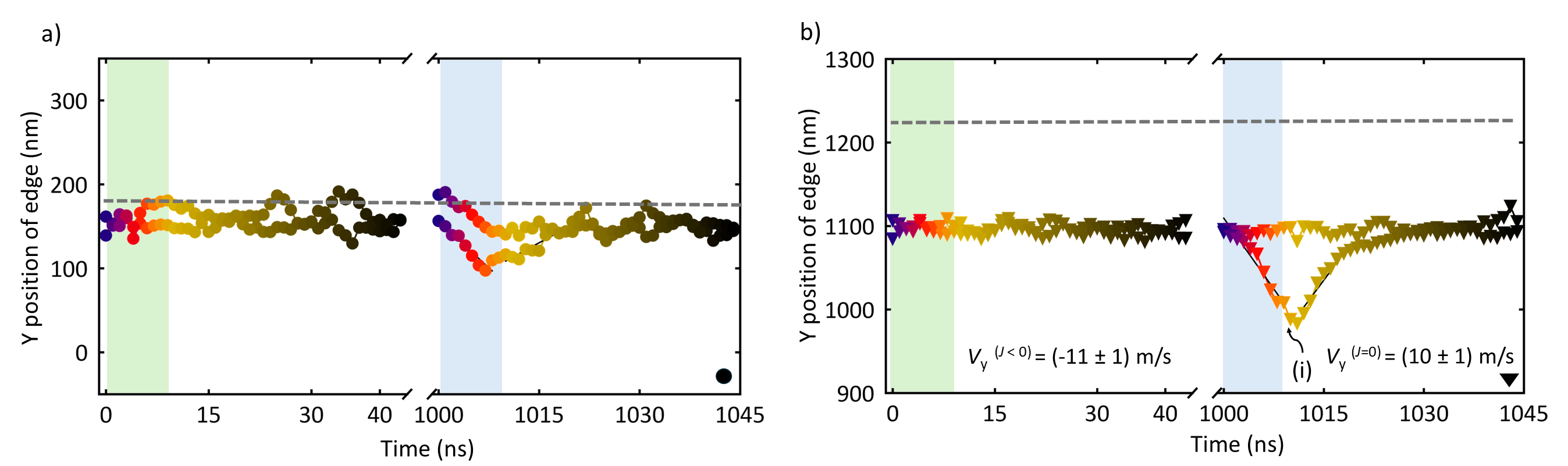}
    \captionsetup{width=0.99\textwidth}
    \caption*{\textbf{Extended Data Fig. 2 | Transverse displacement of the two skymerons shown in Fig. 3}. \textbf{(a,b)}, Time-dependent $y$-positions of the skymerons identified by the circle (a) and triangle (b) markers in Fig. 3a. The green and blue shaded regions indicate the positive- and negative-current pulses, respectively. The colour of the data points encodes the time within each pump--probe cycle. Grey dashed lines indicate the corresponding reference edge positions before excitation. In contrast to the $x$-edge, the current-driven and relaxation velocities in the $y$ direction are comparable, with $v_y^{(J<0)} = (-11 \pm 1)\,\mathrm{m\,s^{-1}}$ and $v_y^{(J=0)} = (10 \pm 1)\,\mathrm{m\,s^{-1}}$.
    }
    \label{figextended2}
\end{figure}

\begin{figure}[h!]
    \centering
    \includegraphics[width = 15cm]{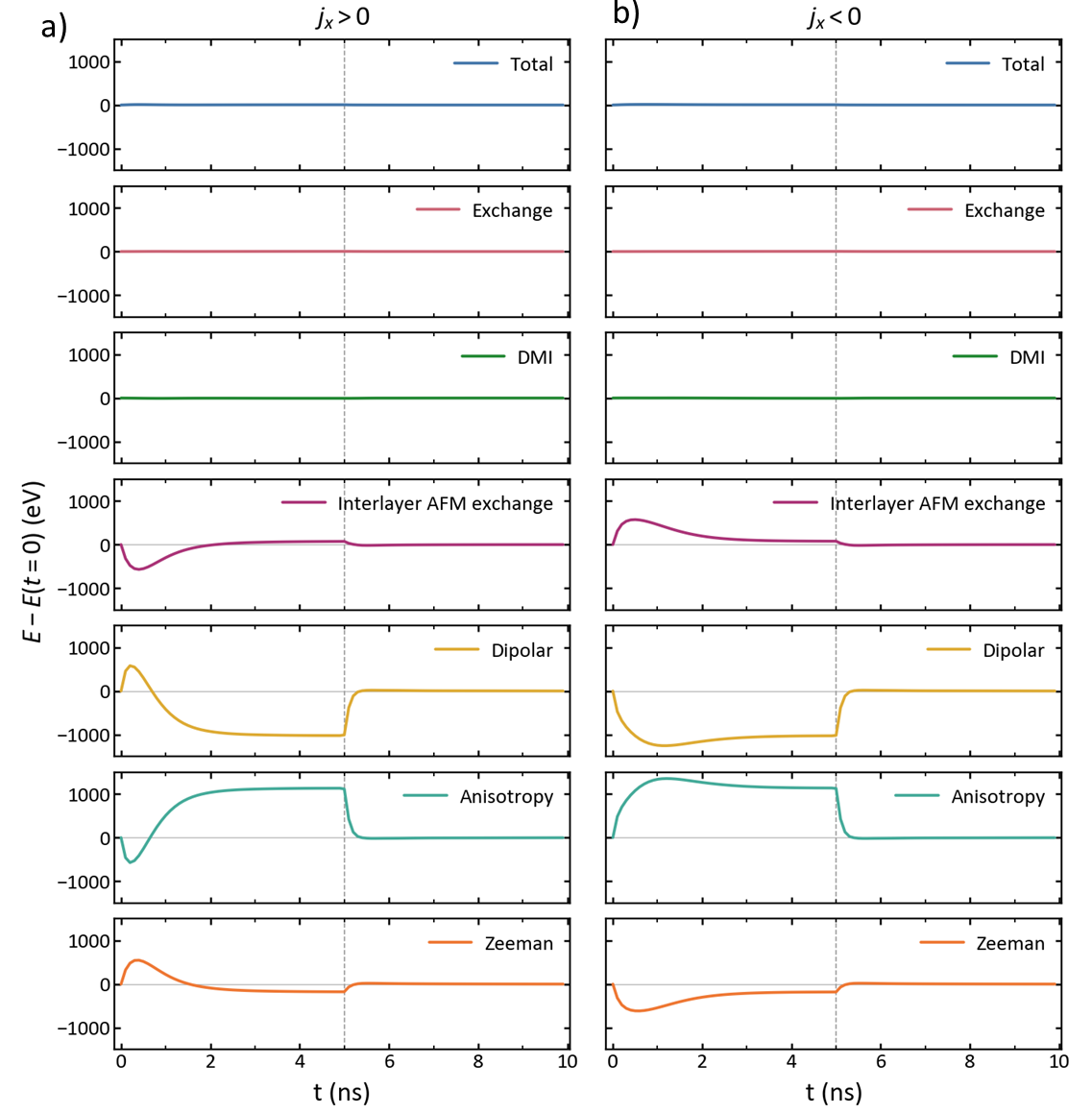}
      \captionsetup{width=0.99\textwidth}

\caption*{\textbf{Extended Data Fig.~3 | Evolution of micromagnetic energy contributions during skymeron dynamics.}
Time evolution of the total, Heisenberg exchange, DMI, interlayer antiferromagnetic exchange, magnetostatic (dipolar), anisotropy, and Zeeman energies for \textbf{a}, $j_x>0$, and \textbf{b}, $j_x<0$, shown relative to their values at $t=0$. The current is applied from $t=0$ to $5~\mathrm{ns}$, with the end of the pulse indicated by the vertical dashed line. The exchange and DMI energies remain essentially unchanged throughout, whereas the dipolar, anisotropy, Zeeman, and interlayer-exchange energies vary substantially as the skymeron expands or contracts depending on current polarity. The anisotropy and magnetostatic contributions vary by approximately$10^{3}~\mathrm{eV}$, with the anisotropy energy developing a pronounced positive excess during the pulse. Once the current is switched off, this excess is redistributed among the other energy terms and relaxes back to its initial value on a finite timescale rather than dissipating instantaneously, defining an effective memory kernel that governs the post-pulse internal dynamics. The delayed release of this stored energy, projected onto the translational coordinate, provides a microscopic basis for the inertia-like motion discussed in the main text.}
        \label{figextended3}
\end{figure}

%\caption*{\textbf{Extended Data Fig.~3 | Evolution of micromagnetic energy contributions during skymeron dynamics.}

%\begin{figure}[h!]
    %\centering
    %\includegraphics[width = 17cm]{contour_skymeron.png}
      %\captionsetup{width=0.99\textwidth}
      %  \caption*{\textbf{Extended Data Figure 2. Contour evolution of representative skymerons}. Time-dependent contours of four representative skymerons extracted from pump--probe XMCD--STXM measurements. The color scale encodes time.1--10\,ns and 46--55\,ns mark the intervals of positive and negative current polarity, respectively. Contours outside these intervals correspond to the evolution after the current pulse is switched off and therefore capture the relaxation dynamics of the texture. Dashed lines indicate the initial skymeron position.
       % }
       % \label{figextended2}
%\end{figure}

\end{document}